\documentclass[letterpaper,12pt]{article}   
\usepackage{osajnl2} 
\usepackage[draft]{hyperref} 

\usepackage{cite}
\usepackage{amsmath,epsfig,multirow}
\usepackage{epstopdf,color,booktabs,comment}
\usepackage[caption=false,font=footnotesize]{subfig}
\usepackage{algorithm}
\usepackage{algorithmic}
\usepackage{amsmath,amsfonts,amsthm,amssymb}
\usepackage{cleveref}

\usepackage{steinmetz}

\DeclareMathOperator*{\argmax}{arg\,max}
\DeclareMathOperator*{\argmin}{arg\,min}

\newcommand{\ak}[1]{\underline{a}^{+}_{1,(#1)}}

\newcommand{\aopt}[0]{\underline{a}_{\sf opt}}
\newcommand{\afoc}[0]{\underline{a}_{\sf foc}}

\newcommand{\focp}[0]{\underline{f}}
\newcommand{\tnorm}[0]{\tau_{\sf normal}}
\newcommand{\topt}[0]{\tau_{\sf opt}}

\newcommand{\ie}{i.e., }
\newcommand{\etal}{et al. }

\newcommand{\aleftp}[0]{\underline{a}^{+}_{1}}
\newcommand{\aleftm}[0]{\underline{a}^{-}_{1}}
\newcommand{\arightp}[0]{\underline{a}^{+}_{2}}
\newcommand{\arightm}[0]{\underline{a}^{-}_{2}}
\newcommand{\flipud}{{\sf flipud}}

\newcommand{\dleftp}[0]{\underline{d}^{+}_{1}}
\newcommand{\dleftm}[0]{\underline{d}^{-}_{1}}

\begin{document}

\title{Iterative, backscatter-analysis algorithms for increasing transmission and focusing light through a highly-scattering random media}


\author{Curtis Jin$^{*}$, Raj Rao Nadakuditi, Eric Michielssen and Stephen C. Rand}
\address{Dept. of EECS, University of Michigan, Ann Arbor, Michigan 48109-2122, USA}
\address{$^*$Corresponding author: jsirius@umich.edu}

\begin{abstract}
Scattering hinders the passage of light through random media and consequently limits the usefulness of optical techniques for sensing and imaging. Thus, methods for increasing the transmission of light through such random media are of interest.
Against this backdrop, recent theoretical and experimental advances have suggested the existence of a few highly transmitting  eigen-wavefronts with transmission coefficients close to one in strongly backscattering random media.

Here, we numerically analyze this phenomenon in 2-D with fully spectrally accurate simulators and provide rigorous numerical evidence confirming the existence of these highly transmitting eigen-wavefronts in random media with periodic boundary conditions that is composed of hundreds of thousands of non-absorbing scatterers.

Motivated by bio-imaging applications where it is not possible to measure the transmitted fields,  we develop physically realizable algorithms for increasing the transmission through such random media using backscatter analysis. We show via numerical simulations that the algorithms converge rapidly, yielding a near-optimum wavefront in just a few iterations. We also develop an algorithm that combines the knowledge of these highly transmitting eigen-wavefronts obtained from backscatter analysis, with intensity measurements at a point to produce a near-optimal focus with significantly fewer measurements than a method that does not utilize this information.

\end{abstract}


\maketitle 

\section{Introduction}
\label{sec:intro}

Media such as glass and air are transparent because light propagates through them without being scattered or absorbed. In contrast, materials such as turbid water, white paint, and egg shells are opaque because the randomly arranged particles cause light to scatter in random directions, thereby hindering its passage. As the thickness of a slab of highly scattering random medium increases, this effect becomes more pronounced, and less and less of a normally incident light is transmitted through \cite{ishimaru1999wave}.

In this context, the theoretical work  of Dorokhov \cite{dorokhov1982transmission}, Pendry \cite{pendry1990maximal,barnes1991multiple}, and others   \cite{mello1988macroscopic,beenakker2009applications} provides unexpected insight into how, and the extent to which, the limitations imposed by random scattering may be overcome. Specifically, these authors  predict that in highly scattering random media composed of non-absorbing scatterers, the eigen-wavefronts associated with the right singular vectors of the $S_{21}$ or transmission matrix will have transmission coefficients whose distribution has a bimodal shape as in Fig.  \ref{fig:dpmk}. Consequently, while many eigen-wavefronts have  a small transmission coefficient,  a small number of eigen-wavefronts exist that have a transmission coefficient close to one, \ie they propagate with almost no scattering loss.

The breakthrough experiments of Vellekoop and Mosk \cite{vellekoop2008phase,vellekoop2008universal} provide evidence of the existence of these highly transmitting eigen-wavefronts in random media.  Vellekoop and Mosk showed \cite{vellekoop2008phase} that intensity measurements on the transmission side of a scattering medium could be used to construct a wavefront that produced about $1000\times$ intensity enhancement at  a target point over that due to a normally incident wavefront. Their work set off a flurry of research on  methods for measuring the transmission matrix and comparing the transmission coefficient distribution with the theoretical prediction \cite{popoff2010measuring,kohlgraf2010transmission,shi2010measuring,kim2012maximal}, faster experimental methods for focusing  \cite{van2011optimal,aulbach2011control,cui2011high,cui2011parallel,stockbridge2012focusing}, and numerical work on the properties of the eigen-wavefronts \cite{choi2011transmission}.

Our work is inspired by these three lines of inquiry. We develop iterative, physically realizable algorithms for transmission maximization that utilize backscatter analysis to produce a highly transmitting wavefront in just a few iterations. These algorithms build on the initial work presented in \cite{jin2012iterative}.

These algorithms which utilize the information in the backscatter field can be useful in applications, such as in bio-imaging, where it might not be possible to measure the transmitted fields.  Our algorithms yield a highly-transmitting wavefront using significantly fewer measurements than required to measure the whole reflection or $S_{11}$ matrix and then generate the wavefront (associated with the smallest right singular vector of the $S_{11}$ matrix) that produces the smallest backscatter (and hence the highest transmission in a lossless medium).

Since our methods maximize transmission by minimizing backscatter, it is important for most of the backscatter field to be captured to fully realize these advantages. Otherwise, given a limited viewing aperture, the principle of backscatter minimization cannot guarantee increased forward transmission and might even produce `transmission' into the unobserved portion of the backscatter field.

 Furthermore, we develop an iterative, physically realizable algorithm for focusing that utilizes intensity measurements at the desired point and backscatter analysis to produce  a near-optimal focusing wavefront with significantly fewer measurements than other approaches. Thus the principal advantage of this approach is that one can get $95$\% of the optimal intensity using significantly fewer measurements than it would take to get the optimal intensity. In effect, we are increasing the rate of convergence to the optimal focusing wavefront. Changing the focusing point or the number of foci do not affect the convergence behavior. We show that we retain this property even when we control fewer than the total number of propagating modes.

A crucial feature of the algorithms we have developed is that it allows the number of modes being controlled via a spatial light modular (SLM) in experiments to be increased without increasing the number of measurements that have to be made.

An additional advantage conferred by these rapidly converging algorithms is that they might facilitate their use in applications  where the duration in which the $S_{21}$ or $S_{11}$ matrix can be assumed to be quasi-static is relatively small compared to the time it would take to make all measurements needed to estimate the $S_{21}$ or $S_{11}$ matrix or in settings where a near-optimal solution obtained fast is  preferable to the optimal solution that takes many more measurements to compute.

Finally, we numerically analyze the phenomenon using a spectrally accurate simulator for 2D scattering systems with periodic boundary conditions  and provide the first numerically rigorous confirmation of the shape of the transmission coefficient distribution and the existence \cite{vellekoop2008universal} of an eigen-wavefront with transmission coefficient approaching one for random media with a large number of scatterers.

The paper is organized as follows. We describe our setup in Section \ref{sec:Setup}. We discuss the problem of transmission maximization and focusing in Section \ref{sec:Prob Form}. To assist in the development of physically realizable algorithms for these applications, we identify physically realizable operations in Section \ref{sec:physical realizable}, and describe iterative, implementable algorithms for finding transmission-maximizing and focusing inputs in Sections \ref{sec:AlgoBackMin} and \ref{sec:FocusingAlgorithms}, respectively. We highlight the existence of the eigen-wavefronts with transmission coefficients approach one, the algorithms' performance and rapid convergence via numerical simulations in Section \ref{sec:simulations}, and summarize our findings in Section \ref{sec:conclusions}.

\section{Setup}
\label{sec:Setup}
\begin{figure}[B]
  \centering
  \includegraphics[trim = 0 18cm 13cm 9cm, clip=on]{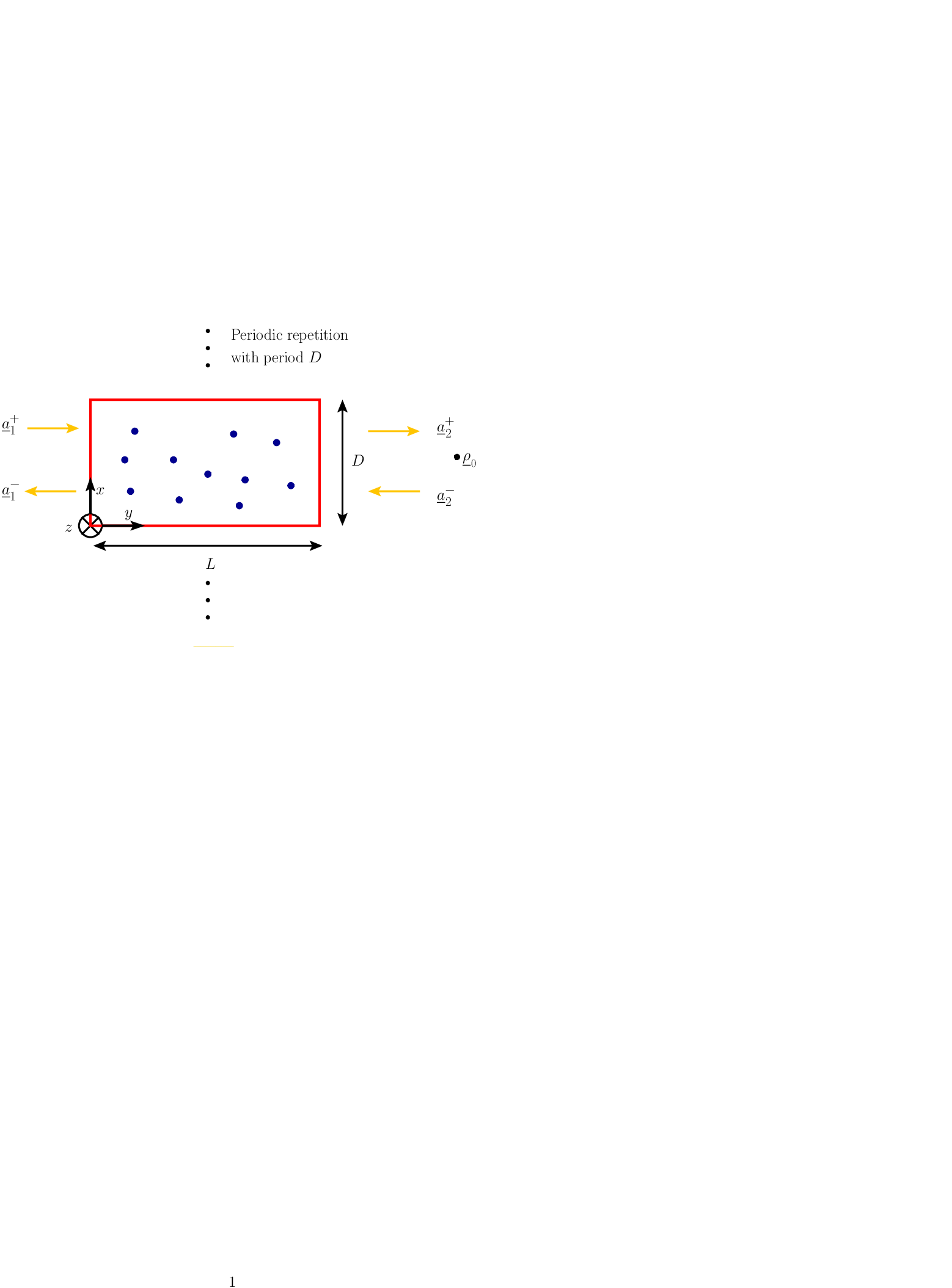}
  \caption{Geometry of the scattering system considered.}
  \label{fig:random model}
\end{figure}
We study scattering from a two-dimensional (2D) random slab of thickness $L$ and periodicity $D$; the slab's unit cell occupies the space $0 \leq x < D$ and $0 \leq y < L$ (Fig.  \ref{fig:random model}). The slab contains $N_{c}$ infinite and $z$-invariant circular cylinders of radius $r$ that are placed randomly within the cell and assumed either perfect electrically conducting (PEC) or dielectric with refractive index $n_{d}$; care is taken to ensure the cylinders do not overlap. Fields are $\sf TM_{\mbox{$z$}}$ polarized: electric fields in the $y<0$ $(i=1)$ and $y>L$ $(i=2)$ halfspaces are denoted $\underline{e}_{i}(\underline{\rho})=e_{i}(\underline{\rho})\hat{z}$. The field (complex) amplitude $e_{i}(\underline{\rho})$ can be decomposed in terms of $+y$ and $-y$ propagating waves as $e_{i}(\underline{\rho}) = e_{i}^{+}(\underline{\rho}) + e_{i}^{-}(\underline{\rho})$, where
\begin{equation}
 \label{eq:IncidentWave}
e^{\pm}_{i}(\underline{\rho}) = \displaystyle \sum_{n=-N}^{N} h_{n} a^{\pm}_{i,n} e^{-j\underline{k}^{\pm}_{n} \cdot \underline{\rho}}\,.
\end{equation}
In the above expression, $\underline{\rho}=x\hat{x}+y\hat{y}\equiv(x,y)$, $\underline{k}^{\pm}_{n} = k_{n,x}\hat{x} \pm k_{n,y}\hat{y} \equiv (k_{n,x},\pm k_{n,y})$, $k_{n,x}=2\pi n/D$, $k_{n,y} = 2\pi \sqrt{(1/\lambda)^{2} - (n/D)^{2}}$, $\lambda$ is the wavelength, and $h_{n}=\sqrt{\| \underline{k}^{\pm}_{n} \|_{2} / k_{n,y}}$ is a power-normalizing coefficient. We assume $N=\lfloor D/\lambda \rfloor$, \ie we only model propagating waves and denote $M=2N+1$. The modal coefficients $a^{\pm}_{i,n}$, $i=1,2$; $n=-N,\ldots,N$ are related by the scattering matrix
\begin{equation}\label{eq:scat matrix}
\left[\begin{array}{c}\aleftm\\\arightp\\\end{array}\right] = \underbrace{\left[ \begin{array}{cc} S_{11} & S_{12} \\ S_{21} & S_{22} \end{array} \right]}_{=:S} \left[\begin{array}{c}\aleftp\\\arightm\\\end{array}\right],
\end{equation}
where $\underline{a}^{\pm}_{i} = \begin{bmatrix} a^{\pm}_{i,-N} & \ldots a^{\pm}_{i,0} & \ldots a^{\pm}_{i,N}\end{bmatrix}^{T}$. In what follows, we assume that the slab is only excited from the $y<0$ halfspace; hence, $\arightm=0$. For a given incident field amplitude $e^{+}_{1}(\underline{\rho})$, we define transmission and reflection coefficients as \begin{align}
\label{eq:tcoeff vec}
\tau(\aleftp) := \dfrac{\| S_{21}\cdot \aleftp \|_{2}^{2}}{\| \aleftp \|_{2}^{2}},\\
\intertext{and}
\label{eq:rcoeff vec}
\Gamma(\aleftp) := \dfrac{\| S_{11}\cdot \aleftp \|_{2}^{2}}{\| \aleftp \|_{2}^{2}},
\end{align}
respectively. We denote the transmission coefficient of a normally incident wavefront by $\tnorm = \tau( \begin{bmatrix} 0 & \cdots &0 & 1 & 0 & \cdots &0 \end{bmatrix}^{T} )$; here $^{T}$ denotes transposition.

\section{Problem formulation}\label{sec:Prob Form}

\subsection{Transmission maximization}

The problem of designing an incident wavefront $\aopt$ that maximizes the transmitted power can be stated as
\begin{equation}\label{eq:optimization problem}
\aopt= \argmax_{\aleftp}  \tau(\aleftp) = \argmax_{\aleftp} \dfrac{\| S_{21}\cdot \aleftp \|_{2}^{2}}{\| \aleftp \|_{2}^{2}} =  \argmax_{\parallel \aleftp\parallel_{2} = 1} \| S_{21} \cdot \aleftp \|_{2}^{2}
\end{equation}
where $\parallel \aleftp \parallel_{2} = 1$ represents the incident power constraint.

Let $S_{21}= \sum_{i=1}^{M} \sigma_{i}\, \underline{u}_i \cdot \underline{v}_{i}^H$ denote the singular value decomposition (SVD) of $S_{21}$; $\sigma_{i}$ is the singular value associated with the left and right singular vectors $\underline{u}_i$ and $\underline{v}_i$, respectively. By convention, the singular values are arranged so that ${\sigma}_{1} \geq \ldots \geq {\sigma}_{M}$ and $^H$ denotes complex conjugate transpose. A well-known result in matrix analysis \cite{horn1990matrix} states that
\begin{equation}\label{eq:aopt s21}
\aopt = {\underline{v}}_{1}.
\end{equation}
When the optimal wavefront $\aopt$ is excited, the optimal transmitted power is $ \topt :=\tau(\aopt) = \sigma_{1}^{2}$. When  the wavefront associated with the $i$-th right singular vector $\underline{v}_{i}$ is transmitted, the transmitted power is $\tau(\underline{v}_{i}) = \sigma_{i}^{2}$, which we refer to as the transmission coefficient of the $i$-th eigen-wavefront of $S_{21}$.  Analogously, we refer to $\Gamma({\underline{v}}_{i})$ as the reflection coefficient of the $i$-th eigen-wavefront of $S_{21}$.

The theoretical distribution \cite{dorokhov1982transmission,pendry1990maximal,barnes1991multiple,mello1988macroscopic,beenakker2009applications}
 of the transmission coefficients for lossless random media has density given by \begin{equation}\label{eq:dpmk}
f(\tau) = \lim_{M \to \infty} \dfrac{1}{M} \sum_{i=1}^{M} \delta\left(\tau-\tau(\underline{v}_{i})\right) =  \dfrac{l}{2L} \dfrac{1}{\tau \sqrt{1-\tau}}, \qquad \textrm{ for } 4 \exp(-L/2l) \lessapprox \tau \leq 1.
\end{equation}
In Eq. (\ref{eq:dpmk}), $l$ is the mean-free path through the medium.  Fig. \ref{fig:dpmk} shows the theoretical density when $L/l= 3$. From,  Eq. (\ref{eq:dpmk}) we expect $\topt = 1$.

From (\ref{eq:aopt s21}) it follows that the optimal wavefront can be constructed by measuring the $S_{21}$ matrix and computing its SVD. Techniques for measuring the $S_{21}$ matrix have been developed in recent works by Popoff \etal \cite{popoff2010measuring} and others  \cite{kohlgraf2010transmission,shi2010measuring}. Kim \etal experimentally measured the $S_{21}$ matrix and demonstrated improved transmission by using the optimal wavefront in Eq. (\ref{eq:aopt s21}) \cite{kim2012maximal}.

In the lossless setting, the scattering matrix $S$ in Eq. (\ref{eq:scat matrix}) will be unitary, \ie  $S^{H}\cdot S = I$, where $I$ is the identity matrix. Consequently, we have that $S_{11}^{H}\cdot S_{11} + S_{21}^{H}\cdot S_{21} = I$, and  the optimization problem in Eq. (\ref{eq:optimization problem}) can be reformulated as
\begin{equation}\label{eq:backscatter min problem}
{\aopt= \argmax_{\parallel \aleftp \parallel_{2}=1 } \underbrace{(\aleftp)^{H} \cdot S_{21}^{H} \cdot S_{21} \cdot \aleftp}_{= (\aleftp)^H \cdot(I - S_{11}^H \cdot S_{11}) \cdot \aleftp}=  \argmin_{\parallel \aleftp\parallel_{2} = 1} \| S_{11} \cdot \aleftp \|_{2}^{2} = \argmin_{ \aleftp}  \Gamma(\aleftp).}
\end{equation}
In other words, in a lossless medium the backscatter-minimizing wavefront also maximizes transmission.
Let $S_{11}= \sum_{i=1}^{M} \widetilde{\sigma}_{i} \widetilde{\underline{u}}_i \cdot \widetilde{\underline{v}}_{i}^H$ denote the SVD of $S_{11}$; $\widetilde{\sigma}_{i}$ is the singular value associated with the left and right singular vectors $\underline{\widetilde{u}}_{i}$ and $\underline{\widetilde{v}}_{i}$, respectively. Then from \cite{horn1990matrix} it follows that
\begin{equation}\label{eq:aopt}
\aopt = \widetilde{\underline{v}}_{M}.
\end{equation}
When this optimal wavefront is excited and the medium is lossless, $\topt = 1- \Gamma(\aopt) = 1- \widetilde{\sigma}_{M}^{2}=\sigma_{1}^{2}$.  When the wavefront associated with the $i$-th right singular vector $\widetilde{\underline{v}}_{i}$ is excited, the transmitted power is given by $\tau(\widetilde{\underline{v}}_{i}) = 1- \Gamma(\widetilde{\underline{v}}_{i}) = 1 - \widetilde{\sigma}_{i}^{2}$, which we refer to as the transmission coefficient of the $i$-th eigen-wavefront of $S_{11}$. Analogously, we refer to $\Gamma(\widetilde{\underline{v}}_{i})$ as the reflection coefficient of the $i$-th eigen-wavefront of $S_{11}$.

A technique for increasing transmission via backscatter analysis would require measurement of the $S_{11}$  matrix and the computation of $\aopt$ as in Eq. (\ref{eq:aopt}). Our objective is to develop fast, physically realizable, iterative algorithms that converge to $\aopt$ by utilizing significantly fewer backscatter field measurements than the $O(M)$ measurements it would take to first estimate $S_{11}$  and then compute its SVD to determine $\widetilde{\underline{v}}_{M}$. Here, we are motivated by applications where it is not possible to measure the transmitted field so that it will not be feasible to measure the $S_{21}$ matrix and compute the optimal wavefront as in Eq. (\ref{eq:aopt s21}).

\subsection{Focusing}
\label{subsec:Focusing}

From Eq. (\ref{eq:IncidentWave}) and using the fact that that $\underline{a}_{2}^{+} = S_{21} \cdot \underline{a}_{1}^{+}$ (since $\underline{a}_{2}^{-} =0$), the field at point $\underline{\rho}_{0}$ is
\begin{equation}
e^{+}_{2}(\underline{\rho}_{0})  = \underbrace{\begin{bmatrix} h_{-N}e^{-j \underline{k}^{+}_{-N} \cdot \underline{\rho}_{0}} & \cdots & h_{N}e^{-j \underline{k}^{+}_{N} \cdot \underline{\rho}_{0}} \end{bmatrix}}_{=: \underline{f}(\underline{\rho}_0)^{H}} \cdot  S_{21} \cdot \aleftp.
\end{equation}
The problem of designing an incident wavefront that maximizes the the intensity (or amplitude squared) of the field at $\underline{\rho}_{0}$  is equivalent to the problem
\begin{equation}\label{eq:focusing problem}
\afoc=\argmax_{\aleftp} \dfrac{ ||e^{+}_{2}(\underline{\rho}_{0})||_{2}^{2}}{||\aleftp ||^{2}_{2}} = \argmax_{\parallel \aleftp\parallel_{2} = 1} \| \underbrace{\focp^{H}(\underline{\rho}_{0}) \cdot S_{21}}_{=:\underline{c}(\underline{\rho}_{0})^{H}} \cdot \aleftp \|_{2}^{2},
\end{equation}
whose solution  is
\begin{equation}\label{eq:afoc}
{\afoc = \dfrac{\underline{c}(\underline{\rho}_{0})}{||\underline{c}(\underline{\rho}_{0})||_{2}} = \dfrac{S_{21}^{H} \cdot \underline{f}(\underline{\rho}_{0})}{|| S_{21}^{H} \cdot \underline{f}(\underline{\rho}_{0})||_{2}}.}
\end{equation}

Thus the optimal  wavefront equals the vector $\underline{c}(\underline{\rho}_{0})$ with normalization to satisfy the power constraint. It can be shown that this wavefront may be obtained by time-reversing the wavefront received by placing a source at $\underline{\rho}_{0}$ \cite{fink1999time}. This fact was exploited in recent work by Cui and collaborators \cite{cui2010implementation,cui2010vivo}.

 In Vellekoop and Mosk's breakthrough work \cite{vellekoop2008phase,vellekoop2008universal,vellekoop2010exploiting}, a coordinate descent method was employed for constructing the optimal wavefront. The coordinate descent approach finds the amplitude and phase of a single mode that maximize the intensity at $\underline{\rho}_0$ while keeping the amplitudes and phases of the other modes fixed and then repeating this procedure for the remaining modes, one mode at a time. In Vellekoop and Mosk's experiments \cite{vellekoop2008phase,vellekoop2008universal,vellekoop2010exploiting}, they kept the amplitude constant for all the modes and considered phase-only modifications of the incident wavefront. While this reduces the complexity of the algorithm, this approach still requires $O(M)$ intensity measurements at $\underline{\rho}_{0}$ to construct the optimal wavefront.  When $M$ is large, the time for convergence will also be large.

This has motivated recent work \cite{cui2011high,cui2011parallel,stockbridge2012focusing} for faster determination of the optimal wavefront. Cui \cite{cui2011high,cui2011parallel} considers an approach using multiple frequencies to find the optimal phases of modes simultaneously, while Stockbridge \etal \cite{stockbridge2012focusing} have proposed a  coordinate descent approach using 2D Walsh functions as a basis set.  These methods have accelerated the experimental convergence, but the reported results are still for small $M$ (between 441 and 1024).

Expressing the optimal wavefront in terms of the singular vectors of $S_{21}$ yields the expression
\begin{equation}
  \label{eq:AFocDecomp}
  \afoc \propto S_{21}^{H} \cdot \underline{f}(\underline{\rho}_{0}) = \sum_{i=1}^{M} \sigma_{i} \underbrace{( \underline{v}_{i}^{H} \cdot \underline{f}(\underline{\rho}_{0}) )}_{=: w_i} \underline{u}_{i} = \sum_{i=1}^{M}\sigma_i w_i \underline{u}_{i}.
\end{equation}

\begin{figure}[t]
 \centering
 \includegraphics[trim = 0 0 0 0, clip = true, width=3.5in]{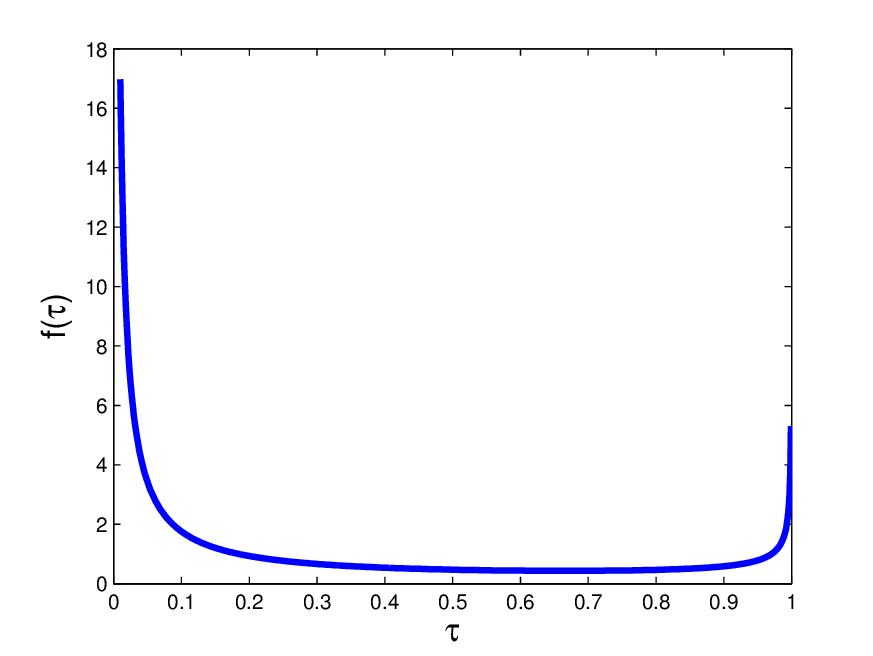}
 \caption{Theoretical distribution in (\ref{eq:dpmk}) for $L/l=3$.}\vspace{-0.35cm}
 \label{fig:dpmk}
\end{figure}

Recall that $\sigma^{2}_{i} = \tau(\underline{v}_{i})$; thus an important insight from Eq. (\ref{eq:dpmk}) and Fig. \ref{fig:dpmk}  is that most of the singular values in Eq. (\ref{eq:AFocDecomp}) are close to zero. However, there typically are $ K \ll M$ singular values close to one. It is the superposition of these $K$ eigen-wavefronts of $S_{21}$ having transmission coefficients close to one whose constructive interference yields the maximal transmission that contributes to maximal intensity.


In the lossless setting, when the scattering matrix $S$  is unitary, we have that $\tau(\underline{v}_{i}) = 1 - \Gamma(\widetilde{\underline{v}}_{M-i+1})$. Hence, the $K$ eigen-wavefronts of $S_{21}$ that have transmission coefficients close to one correspond precisely to the $K$ eigen-wavefronts associated with $S_{11}$ that have reflection coefficients close to zero.  By using $O(K)$ backscatter field measurements to measure the $K$ eigen-wavefronts of $S_{11}$ with small reflection coefficients and $O(K)$ intensity measurements at $\underline{\rho}_{0}$, we might expect to approximate $\afoc$ in Eq.  (\ref{eq:AFocDecomp}) and yield a near-optimal focus using just $O(K)$ measurements (we expect $K \ll M$).

Our objective is to develop a fast, physically realizable, iterative algorithm that utilizes backscatter field measurements and intensity measurements at $\underline{\rho}_{0}$ to construct a near-optimal focusing wavefront using significantly fewer measurements than are required by coordinate descent methods that only employ intensity measurements at $\underline{\rho}_{0}$. The emphasis here is on accelerating the convergence behavior; we do not improve the quality of the focus.

\section{Recognizing physically realizable matrix-vector operations}
\label{sec:physical realizable}
The iterative algorithms we will develop in Sections \ref{sec:AlgoBackMin} and
\ref{sec:FocusingAlgorithms} build on the vast literature of iterative methods in numerical linear algebra \cite{van2003iterative,trefethen1997numerical}. The algorithms are based on three matrix-vector operations, $S_{11}\cdot \aleftp$, $F \cdot (\aleftm)^{*}$, and $S_{11}^{H}\cdot \aleftm$. These operations can be performed mathematically, but the measurement corresponding to these operations in a physical setting is not obvious. Here, we dwell on mapping these matrix-vector operations into their physical counterparts, thus making our algorithms physically realizable.

{ The first operation, $S_{11}\cdot \aleftp$, can be realized by measuring the backscattered wave.} In an experimental setting, the modal coefficient vector of the backscattered wave would be extracted from the backscatter intensity measurement by digital holography techniques described in, for example \cite{grilli2001whole}. We also assume that it is possible to modulate the amplitude and phase of a wavefront, using the methods described in \cite{van2008spatial}. Thus, the matrix-vector multiplicative operation $S_{11}\cdot \aleftp$ corresponds to sending an incident wavefront with modal coefficient vector $\aleftp$ and measuring the modal coefficient vector of the backscattered wavefront. Furthermore, we assume that these modal coefficient vectors can be recovered perfectly, and the amplitude and the phase can be perfectly modulated, so that we might investigate the best-case performance of the algorithms.

{ The second operation, $F\cdot (\aleftm)^{*}$, can be realized by time-reversing the wave.} Let $\flipud(\cdot)$ represent the operation of flipping a vector or a matrix argument upside down so that the first row becomes the last row and so on, and let $^*$ denote complex conjugation. We define $F = \flipud(I)$, where $I$ is the identity matrix; then the operation $F \cdot (\aleftm)^{*}$ represents time-reversing the wave corresponding to $\aleftm$. This can be explained as follows. The expression for time-reversed wave of $\aleftm$ is \begin{align}
  \nonumber (e_{1}^{-}(\underline{\rho}))^{*} = \left( \displaystyle \sum_{n=-N}^{N} h_{n} a^{-}_{1,n} e^{-j\underline{k}^{-}_{n} \cdot \underline{\rho}} \right)^{*} &= \displaystyle \sum_{n=-N}^{N} h_{n}^{*} (a^{-}_{1,n})^{*} e^{j\underline{k}^{-}_{n} \cdot \underline{\rho}} \\
  \label{eq:TRSModalVector} &= \displaystyle \sum_{n=-N}^{N} h_{n} (a^{-}_{1,-n})^{*} e^{-j\underline{k}^{+}_{n} \cdot \underline{\rho}}.
\end{align}
Note that we have used the fact that $h_{-n}^{*} = h_{n}$ and $\underline{k}^{-}_{-n} = -\underline{k}^{+}_{n}$. From Eq.  (\ref{eq:TRSModalVector}), we see that the modal coefficient vector representation of the time-reversed wave of $\aleftm$ is $\begin{bmatrix} (a_{N}^{-})^{*} & (a_{N-1}^{-})^{*} & \ldots & (a_{-N+1}^{-})^{*} & (a_{-N}^{-})^{*} \end{bmatrix}^{T}=F\cdot (\aleftm)^{*}$. Furthermore, we emphasize that the operation $F \cdot (\aleftm)^{*}$ can be physically realized via phase-conjugate mirroring (PCM) \cite{fink1999time}.

\begin{figure}[h]
\begin{center}
\vspace{-0.35cm}
\includegraphics[trim = 0 23.5cm 11cm 3cm, clip=on]{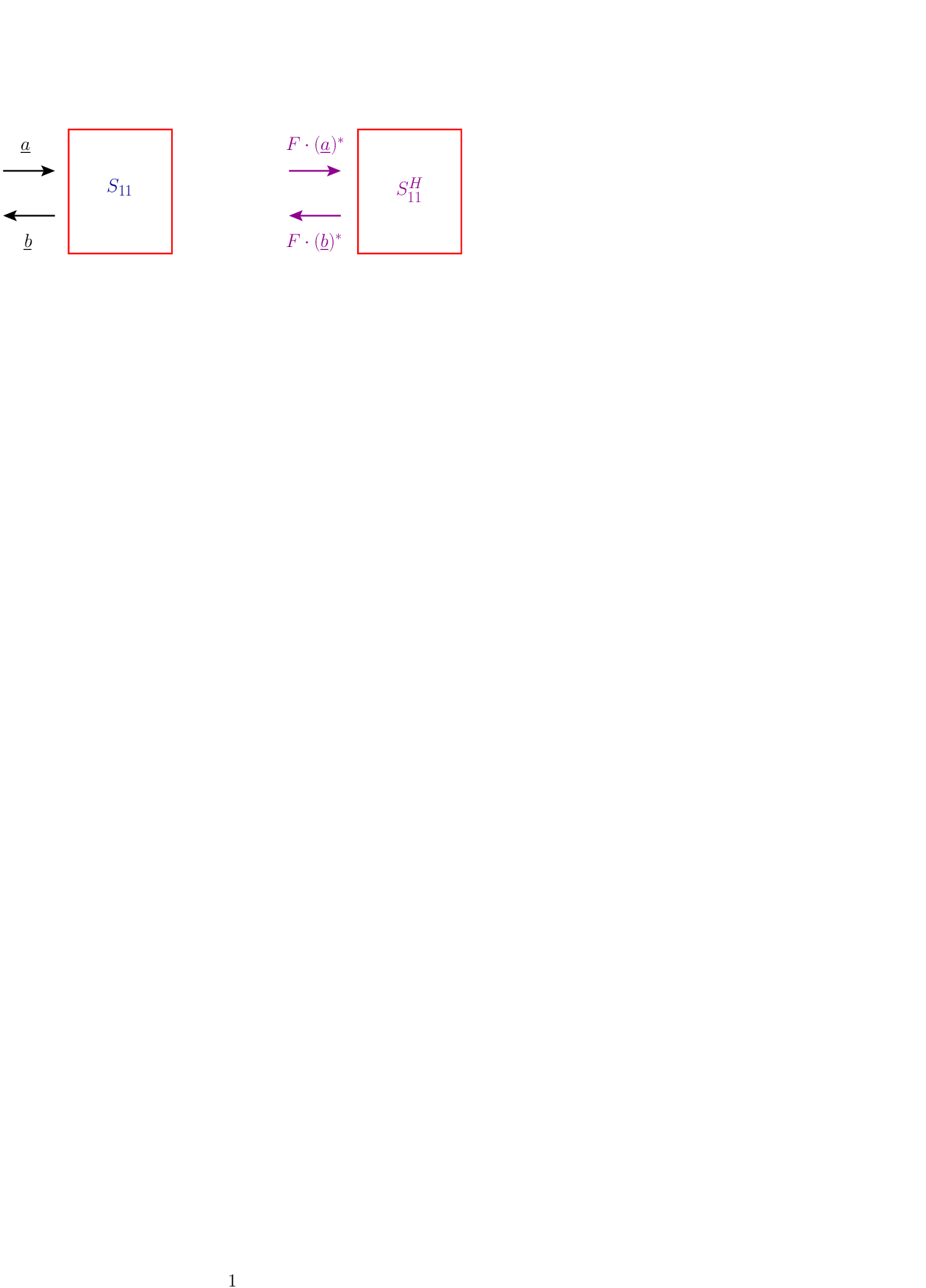}
\vspace{-0.35cm}
\caption{The relationship between wavefronts in a medium that exhibits reciprocity. Reciprocity tells us that $S_{11}^{H}\cdot \underline{a}$ is obtained by time-reversing the wave \emph{before and after} sending $\underline{a}$ into the medium, and we call this sequence of operations \emph{double phase conjugation}.}
\label{fig:Reciprocity}
\vspace{-0.35cm}
\end{center}
\end{figure}

{The third operation, $S_{11}^{H}\cdot \aleftm$, can be realized by using reciprocity.} In a scattering medium that exhibits reciprocity, there are relationships \cite{kong1986electromagnetic,haus1984waves,carminati2000reciprocity,potton2004reciprocity,nieto1991scattering} between the incident and scattered wavefronts. Consequently, reciprocity requires the reflection matrix $S_{11}$  to satisfy
\begin{equation}\label{eq:S11 op}
S_{11}^{H} = F\cdot S_{11}^{*} \cdot F.
\end{equation}
This means that if $\underline{a}$ is an input to the system that produces a backscattered wave of $\underline{b}$, then sending $F\cdot (\underline{a})^{*}$ will produce backscattered wave of $F\cdot (\underline{b})^{*}$ in a medium whose reflection matrix corresponds to $S_{11}^{H}$. (Fig.  \ref{fig:Reciprocity})

An important implication of this equation is that the matrix-vector operation $S_{11}^{H}\cdot \aleftm$ can be cast in terms of physically realizable operations. Note that $S_{11}^{H}\cdot \aleftm$  can be expressed as
\begin{equation}
\nonumber S_{11}^{H}\cdot \aleftm = F\cdot S_{11}^{*} \cdot F \cdot \aleftm = F \cdot (S_{11} \cdot (F\cdot (\aleftm)^{*}))^{*}.
\end{equation}
\noindent From the last expression, we see that the operation $S_{11}^{H} \cdot \aleftm$ can be physically realized in a sequence of two steps:
\begin{enumerate}
\item{Time-reverse the wavefront whose modal coefficient vector is $\aleftm$, and send it to the scattering system.}
\item{Time-reverse the resulting backscattered wavefront.}
\end{enumerate}
We call this sequence of operations as \emph{double phase conjugation}, and we shall leverage it extensively in what follows.

\setlength{\heavyrulewidth}{0.1em}
\newcommand{\otoprule}{\midrule[\heavyrulewidth]}
\renewcommand*\arraystretch{1.5}

\section{Iterative, physically realizable algorithms for transmission maximization}
\label{sec:AlgoBackMin}
We now develop iterative, physically realizable algorithms for transmission maximization that converge to $\aopt$ in Eq. (\ref{eq:aopt}), by utilizing significantly fewer backscatter field measurements than the $O(M)$ measurements it would take to first estimate $S_{11}$  and then compute its SVD to determine $\widetilde{\underline{v}}_{M}$.

\subsection{Steepest descent method}
\label{subsec:SD}

The backscatter minimization problem involves optimization with respect to the objective function $\|S_{11}\cdot \aleftp \|_{2}^{2}$ that appears on the right hand side of Eq. (\ref{eq:backscatter min problem}). The objective function's negative gradient is used as a search direction to correct the previous input as\begin{equation}
\nonumber \ak{k+1} = \ak{k} - \mu \left. \frac{\partial \| S_{11} \cdot \aleftp \|_{2}^{2} }{\partial \aleftp} \right|_{\aleftp = \ak{k}} = \ak{k} -2\mu S_{11}^{H} \cdot S_{11} \cdot \ak{k},
\end{equation} where $\ak{k}$ represents the modal coefficient vector of the wavefront produced at the $k$-th iteration of the algorithm and $\mu$ is a positive stepsize.  This yields Algorithm \ref{alg:SD} which iteratively refines the wavefront $\ak{k+1}$ until the backscattered intensity $\| S_{11} \cdot \ak{k} \|_{2}^{2}$ drops below a preset threshold $\epsilon$.

\begin{algorithm}
\centering
\caption{Steepest descent algorithm for finding \textbf{$\aopt$}}
\label{alg:SD}
\begin{algorithmic}[1]\label{alg}
\STATE Input: $\ak{0} = \mbox{ Initial random vector with unit norm}$
\STATE Input: $\mu>0 = \mbox{step size}$
\STATE Input: $\epsilon = $ Termination condition
\STATE $k=0$
\WHILE{$\| S_{11} \cdot \ak{k} \|_{2}^{2} > \epsilon $}
\STATE $\underline{\tilde{a}}^{+}_{1,(k)} = \ak{k} - 2 \mu S_{11}^{H}\cdot S_{11} \cdot \ak{k}$
\STATE $\ak{k+1} = \underline{\tilde{a}}^{+}_{1,(k)} / \| \underline{\tilde{a}}^{+}_{1,(k)} \|_{2}$
\STATE $k = k + 1$
\ENDWHILE
\end{algorithmic}
\end{algorithm}

\indent Armed with the relationship in Eq. (\ref{eq:S11 op}), step 6 in Algorithm \ref{alg:SD} can be expressed as
\begin{equation}
\underline{\tilde{a}}^{+}_{1,(k)} = \ak{k} - 2 \mu S_{11}^{H}\cdot S_{11} \cdot \ak{k} =  \ak{k} - 2 \mu F \cdot S_{11}^{*}\cdot F \cdot S_{11} \cdot \ak{k}.
\end{equation}
This allows us to recast each step of Algorithms \ref{alg:SD}  into the counterparts of the physical operations in the second column of Table \ref{tab:SD}.

\begin{table}[]
    \centering
    \begin{tabular}{ll}
   \otoprule
   \textbf{Vector Operation} & \textbf{Physical Operation}\\
   \hline
    $1: \quad \aleftm=S_{11}\cdot \ak{k}$ & $1: \quad  \ak{k} \xrightarrow{\mbox{ \tiny Backscatter \quad}} \underline{a}_{1}^{\tiny -}$\\
    $2: \quad \aleftp=F \cdot (\aleftm)^{*}$ & $2: \quad  \underline{a}_{1}^{\tiny -} \xrightarrow{\mbox{ \tiny PCM \quad}} \underline{a}_{1}^{\tiny +}$\\
    $3: \quad \aleftm=S_{11} \cdot \aleftp$ & $3: \quad  \aleftp \xrightarrow{\mbox{ \tiny Backscatter \quad}} \underline{a}_{1}^{\tiny -}$\\
    $4: \quad \aleftp=F \cdot (\aleftm)^{*}$ & $4: \quad  \underline{a}_{1}^{\tiny -} \xrightarrow{\mbox{ \tiny PCM \quad}} \underline{a}_{1}^{\tiny +}$\\
    $5: \quad \underline{\tilde{a}}^{+}_{1}=\ak{k}-2\mu \aleftp$ & $5: \quad  \underline{\tilde{a}}^{+}_{1}=\ak{k}-2\mu \aleftp$\\
    $6: \quad \ak{k+1}=\underline{\tilde{a}}^{+}_{1} / \mbox{ } \|\underline{\tilde{a}}_{1}^{+}\|_{2}$ & $6: \quad  \underline{\tilde{a}}_{1}^{\tiny +} \xrightarrow{\mbox{ \tiny Normalization \quad}} \ak{k+1}$\\
  \otoprule
 \end{tabular}
 \caption{Steepest descent algorithm for transmission maximization. The first column represents vector operations in Algorithm \ref{alg:SD}. The second column represents the physical (or experimental) counterpart. The operation $\aleftm \longmapsto F\cdot (\aleftm)^{*}$ can be realized via the use of a phase-conjugating mirror (PCM). The algorithm terminates when the backscatter intensity falls below a preset threshold $\epsilon$.}
\vspace{-0.45cm}
 \label{tab:SD}
\end{table}

The sequence of steps $1-4$ in Table \ref{alg:SD},which involves \emph{double phase conjugation}, amplifies the highly-backscattering component in the wavefront, analogous to the operations for time-reversal focusing \cite{prada1994eigenmodes,fink1989self,fink1999time,yanik2004time}. In step $5$, this component is subtracted leading to a refined wavefront that will backscatter less. This process is repeated till convergence.  A consequence of this technique is that the backscatter field intensity will typically decrease monotonically. This makes the measurement of the backscatter modal coefficient vector increasingly difficult as the iteration progresses. An additional disadvantage of this method is the obvious need to carefully set $\mu$ to guarantee convergence, $0 < \mu < \frac{1}{\widetilde{\sigma}^{2}_{1}+\widetilde{\sigma}^{2}_{M}} \approx 1$. In an experimental setting, the step size $\mu$ is chosen by a simple line search, \ie by scanning a set of discretized values and selecting the one that results in the smallest backscatter intensity after a fixed number of iterations.

We describe a method next, which maintains high backscatter field intensity throughout the process and does not require selection of any other auxiliary parameters to guarantee convergence.

\subsection{Conjugate gradient method}
\label{subsec:CG}

Consider an iterative solution to Eq. (\ref{eq:backscatter min problem}) where the iterate (before normalization for power) is formed as
\begin{equation}\label{eq:cg akp1}
\ak{k+1} = \ak{k} + \mu_{(k+1)}\underline{d}_{(k)},
\end{equation}
where $\mu_{(k+1)}$ is a stepsize and $d_{(k)}$ is the search direction. In this framework, Algorithm \ref{alg:SD} results from setting $\mu_{(k+1)} = \mu$ and $\underline{d}_{(k)} = -2 S_{11}^{H} \cdot S_{11} \cdot \ak{k}$.

The conjugate gradients method (see \cite[Chapter 5]{van2003iterative} for a detailed derivation) results from choosing the stepsize
\begin{subequations}
\begin{equation}\label{eq:cg mukp1}
\mu_{(k+1)} = \|\underline{r}_{(k)}\|_{2}^2 / \| S_{11} \cdot \underline{d}_{(k)} \|_{2}^2,
\end{equation}
\text{with the search direction given by}
\begin{equation}\label{eq:cg dkp1}
\underline{d}_{(k+1)} = \underline{r}_{(k+1)} + \beta_{(k+1)} \underline{d}_{(k)},
\end{equation}
\text{and}
\begin{equation}\label{eq:cg betakp1}
\beta_{(k+1)} = \|\underline{r}_{(k+1)}\|_{2}^2 / \|\underline{r}_{(k)}\|_{2}^2.
\end{equation}
\text{Here, the residual vector is}
\begin{equation}\label{eq:cg rkp1}
\underline{r}_{(k+1)} = -S_{11}^{H}\cdot S_{11} \cdot \ak{k+1}.
\end{equation}
\end{subequations}
The iteration terminates when  $|| \underline{r}_{(k+1)}||_{2} < \epsilon $, a preset threshold.

Plugging Eq. (\ref{eq:cg akp1}) into Eq. (\ref{eq:cg rkp1}) and substituting the expressions in Eqs.(\ref{eq:cg mukp1}) - (\ref{eq:cg betakp1}) gives us an alternate expression for the residual vector
\begin{subequations}
\begin{equation}
\underline{r}_{(k+1)} = \underline{r}_{(k)} - \mu_{(k+1)}S_{11}^{H}\cdot S_{11} \cdot \underline{d}_{(k)},
\end{equation}
\text{or, equivalently}
\begin{equation}\label{eq:cg alt res}
\underline{r}_{(k+1)}= \underline{r}_{(k)} -  \dfrac{\|\underline{r}_{(k)}\|_{2}^2}{\| S_{11} \cdot \underline{d}_{(k)} \|_{2}^2} S_{11}^{H}\cdot S_{11} \cdot \underline{d}_{(k)}.
\end{equation}
\end{subequations}
The utility of Eq. (\ref{eq:cg alt res}) will become apparent shortly.

To summarize: we described an iterative method for refining the wavefront $\ak{k}$ via Eq. (\ref{eq:cg akp1}). Inspection of the update Eqs. (\ref{eq:cg mukp1})-(\ref{eq:cg betakp1}) and Eq. (\ref{eq:cg alt res}) reveals that matrix-vector operations  $S_{11} \cdot \underline{d}_{(k)}$ appears in Eq. (\ref{eq:cg mukp1}) while $S_{11}^{H}\cdot S_{11} \cdot \underline{d}_{(k)}$ appears in Eq. (\ref{eq:cg alt res}). This means that the vector $\underline{d}_{(k)}$ is transmitted and the associated backscatter is measured. Note that these measurements are used to iteratively refine the vector $\ak{k}$ , but $\ak{k}$ is \textit{never actually transmitted} until the termination condition $|| \underline{r}_{(k+1)}||_{2} < \epsilon$  is met. This is reflected in the physical description of the proposed algorithm in Table \ref{tab:CG}. Also, note that we start with a random unit vector $\ak{0}$, and set $\underline{d}_{(0)}$ and $\underline{r}_{(0)}$ to $-S_{11}^{H}\cdot S_{11} \cdot \ak{0}$, since we are using conjugate gradient for finding the input that minimizes reflection, $\ie$ $$ -\ak{0} \xrightarrow{\mbox{ \tiny Backscatter \quad}} \aleftm \xrightarrow{\mbox{ \tiny PCM \quad}} \underline{a}_{1}^{\tiny +} \xrightarrow{\mbox{ \tiny Backscatter \quad}} \underline{a}_{1}^{\tiny -} \xrightarrow{\mbox{ \tiny PCM \quad}} \underline{d}_{(0)} = \underline{r}_{(0)} .$$

\begin{table}[h]
    \centering
    \begin{tabular}{ll}
   \otoprule
   \textbf{Vector Operation} & \textbf{Physical Operation}\\
   \hline
    $1: \quad \dleftm=S_{11}\cdot \underline{d}_{(k)}$ & $1: \quad  \underline{d}_{(k)} \xrightarrow{\mbox{ \tiny Backscatter \quad}} \underline{d}_{1}^{\tiny -}$\\
    $2: \quad \dleftp=F \cdot (\aleftm)^{*}$ & $2: \quad  \underline{d}_{1}^{\tiny -} \xrightarrow{\mbox{ \tiny PCM \quad}} \underline{d}_{1}^{\tiny +}$\\
    $3: \quad \dleftm=S_{11} \cdot \dleftp$ & $3: \quad  \dleftp \xrightarrow{\mbox{ \tiny Backscatter \quad}} \underline{d}_{1}^{\tiny -}$\\
    $4: \quad \underline{d}=F \cdot (\dleftm)^{*}$ & $4: \quad  \underline{d}_{1}^{\tiny -} \xrightarrow{\mbox{ \tiny PCM \quad}} \underline{d}$\\
    $5: \quad \mu_{(k+1)} = \|\underline{r}_{(k)}\|_{2}^{2} /(\underline{d}_{(k)}^{H} \cdot \underline{d} )$  & $5: \quad  \mu_{(k+1)} = \|\underline{r}_{(k)}\|_{2}^{2} /(\underline{d}_{(k)}^{H} \cdot \underline{d} )$\\
    $6: \quad \underline{r}_{(k+1)} = \underline{r}_{(k)} - \mu_{(k+1)} \underline{d}$ & $6: \quad  \underline{r}_{(k+1)} = \underline{r}_{(k)} - \mu_{(k+1)} \underline{d}$\\
    $7: \quad \beta_{(k+1)} = \|\underline{r}_{(k+1)}\|_{2}^{2} / \|\underline{r}_{(k)}\|_{2}^{2}$ & $7: \quad  \beta_{(k+1)} = \|\underline{r}_{(k+1)}\|_{2}^{2} / \|\underline{r}_{(k)}\|_{2}^{2}$\\
    $8: \quad \underline{d}_{(k+1)} = \underline{r}_{(k+1)} + \beta_{(k+1)} \underline{d}_{(k)}$ & $8: \quad  \underline{d}_{(k+1)} = \underline{r}_{(k+1)} + \beta_{(k+1)} \underline{d}_{(k)}$\\
  \otoprule
 \end{tabular}
 \caption{Conjugate gradient algorithm for transmission maximization. The first column represents iterates of the conjugate gradients method. The second column represents the physical (or experimental) counterpart. The operation $\aleftm \longmapsto F\cdot (\aleftm)^{*}$ can be realized via the use of a phase-conjugating mirror (PCM). The algorithm terminates when the residual vector $||\underline{r}_{(k+1)}||_{2} < \epsilon,$ a preset threshold at which point the optimal backscatter minimizing wavefront is constructed as $\ak{k+1} = \ak{k} + \mu_{(k+1)}\underline{d}_{(k)}$ followed by a power normalization $\ak{k+1} = \ak{k+1}/||\ak{k+1}||_2$.}
\vspace{-0.45cm}
 \label{tab:CG}
\end{table}

A feature of the conjugate gradient method is that the intensity of the backscatter measurement $S_{11} \cdot \underline{d}_{(k)}$ is expected to remain relatively high (for a strongly backscattering medium) throughout the process. It is only when the wavefront corresponding to $\ak{k+1}$ is excited that a strong transmission (with minimized backscatter) is obtained - this might be a desirable  feature for communication or covert sensing applications. Consequently, the algorithm will produce high intensity backscatter measurements, thereby facilitating accurate estimation of the backscatter modal coefficient vectors that are an important component of the proposed algorithm. This makes the conjugate gradient method less susceptible to measurement noise than the steepest descent method where the backscatter intensity decreases with every iteration.

\section{An iterative, physically realizable focusing algorithm}
\label{sec:FocusingAlgorithms}

We first describe a generalized coordinate descent method for amplitude and phase optimization. Assume we are given a $M \times N_{B}$ matrix $B= \left[ \underline{b}_{1} \quad \ldots \quad \underline{b}_{N_{B}} \right]$ whose columns are orthonormal so that $B^{H} \cdot B = I_{N_{B}}$. Thus $N_{B}$ denotes the number of (orthonormal) bases vectors.

The key idea here is to expand $\underline{a}_{1}^{+}$ on the right hand side of Eq. (\ref{eq:focusing problem}) in terms of the bases vectors given by the columns of $B$ as
\begin{equation} \label{eq:modal decomp B}
\underline{a}_{1}^{+}= \sum_{l = 1}^{N_{B}} p_{l} e^{j \phi_{l}} \underline{b}_{l},
\end{equation}
where $p_{l} \geq 0$  and $\phi_{l} \in [-\pi, \pi]$ are the unknown amplitudes and phases, respectively.

The optimal amplitudes can be estimated by transmitting $\underline{a}_{1}^{+}=\underline{b}_{l} $ for every $l = 1, \ldots N_{B}$, measuring the corresponding intensity $\mathcal{I}_{l}$ at the target, and setting $p_{l} = \sqrt{\mathcal{I}_{l}}$. This can be accomplished with $O(N_{B})$ measurements.

The phases can be estimated by first setting $\phi_{1}, \ldots \phi_{N_{B}}$ randomly and then for $l = 1, \ldots, N_{B}$, sequentially finding the phase that optimizes measured intensity. This can be done via a simple line search, \ie by scanning the measured intensity over a fixed set of discretized values of the phase or by using more sophisticated algorithms such as golden section search algorithm with parabolic interpolation \cite[Section 10.2]{press2007numerical}. This too requires $O(N_{B})$ measurements.

Setting $N_{B} = M$ and $B = I$ yields the coordinate descent approach used by Vellekoop and Mosk \cite{vellekoop2008phase,vellekoop2008universal,vellekoop2010exploiting}. This corresponds to exciting one plane wave mode at a time and inferring the optimal phase and amplitude one mode at time. Such an algorithm requires $O(M)$ iterations to yield the optimal focussing wavefront.  Setting $B$ to the 2D Walsh function basis matrix yields the method proposed by Stockbridge \etal  in \cite{stockbridge2012focusing}.

An  important insight from Eq. (\ref{eq:AFocDecomp})  is that if we were to express the optimal focusing wavefront as a superposition of eigen-wavefronts of $S_{21}$, then typically only $K \ll M$ of the combining coefficients will be large. Thus only $K$ of the $p_{l}$ coefficients in Eq. (\ref{eq:modal decomp B}) will be significant if we set $B$ to be the right singular vectors of $S_{21}$. In the lossless setting, the $K$ eigen-wavefronts of $S_{21}$ that have transmission coefficients close to one correspond precisely to the $K$ eigen-wavefronts associated with $S_{11}$ that have reflection coefficients close to zero. Hence, we can set $B$ to be the right singular vectors of $S_{11}$ and expect only $K$ of the $p_{l}$ coefficients in Eq. (\ref{eq:modal decomp B}) to be significant as well.  Thus, we need to measure the $K$ singular vectors of $S_{11}$ associated with its $K$ smallest singular values.

The Lanczos algorithm is an iterative algorithm for accomplishing just that \cite{van2003iterative,trefethen1997numerical}. The key idea is to create a tridiagonal matrix $H$ whose eigenvalues and eigenvectors (referred to as the Ritz values and vectors) are approximations of the eigenvalues and eigenvectors of $S_{11}^{H}\cdot S_{11}$. The algorithm is summarized in the first column of Table \ref{tab:ChanIdent}; its physical counterpart is described in the second column. The matrix $B$ in Eq. (\ref{eq:modal decomp B}) is obtained as
\begin{equation}\label{eq:Bfoc}
B = Q \cdot U,
\end{equation}
where $Q = \begin{bmatrix} \underline{q}_{(1)} & \ldots &  \underline{q}_{(N_{B})}\end{bmatrix}$ are the $N_{B}$ vectors produced by the algorithm (see Table \ref{tab:ChanIdent}) and $U = \begin{bmatrix} \underline{u}_{(1)} & \ldots &  \underline{u}_{(N_{B})}\end{bmatrix}$ are the $N_{B}$ eigenvectors of $H$ associated with the $N_{B}$ smallest eigenvalues.

The convergence theory \cite{trefethen1997numerical} of the Lanczos algorithms predicts that the eigenvector estimates will rapidly converge to the $K$ eigenvectors of $S_{11}^{H}\cdot S_{11}$ associated with the eigen-wavefronts of $S_{11}$ with the smallest reflection coefficients; hence,  setting $N_{B} = O(K)$ will suffice. An estimate of $K$ can be formed from the eigenvalues of $H$ by counting how many of the converged eigenvalues of $H$ are below a preset threshold $\epsilon$.

Estimating these $K$ right singular vectors  will require $O(K)$ measurements and when $K \ll M$, we shall obtain a near-optimal focusing wavefront using significantly fewer measurements  than the $O(M)$ measurements required by the coordinate descent when $B = I$. We shall corroborate this convergence behavior using numerical simulations next.

\begin{table}[t]
    \centering
    \begin{tabular}{ll}
   \otoprule
   \textbf{Vector Operation} & \textbf{Physical Operation}\\
    \hline
    $1: \quad \underline{q}_{1}^{-}=S_{11}\cdot \underline{q}_{(k)}$ & $1: \quad  \underline{q}_{(k)} \xrightarrow{\mbox{ \tiny Backscatter \quad}} \underline{q}_{1}^{-}$\\
    $2: \quad \underline{q}_{1}^{+}=F \cdot (\underline{q}_{1}^{-})^{*}$ & $2: \quad  \underline{q}_{1}^{-} \xrightarrow{\mbox{ \tiny PCM \quad}} \underline{q}_{1}^{+}$\\
    $3: \quad \underline{q}_{1}^{-}=S_{11} \cdot \underline{q}_{1}^{+}$ & $3: \quad  \underline{q}_{1}^{+} \xrightarrow{\mbox{ \tiny Backscatter \quad}} \underline{q}_{1}^{-}$\\
    $4: \quad \underline{v}=F \cdot (\underline{q}_{1}^{-})^{*}$ & $4: \quad  \underline{q}_{1}^{-} \xrightarrow{\mbox{ \tiny PCM \quad}} \underline{v}$\\
    $5: \quad H_{k,k} = \underline{q}_{(k)}^{H} \cdot \underline{v}$ & $5: \quad H_{k,k} = \underline{q}_{(k)}^{H} \cdot \underline{v}$\\
    $6: \quad \underline{v} = \underline{v} - H_{k,k}\underline{q}_{(k)} - s_{(k-1)} \underline{q}_{(k-1)}$ & $6: \quad \underline{v} = \underline{v} - H_{k,k}\underline{q}_{(k)} - s_{(k-1)} \underline{q}_{(k-1)}$\\
    $7: \quad H_{k+1,k} = H_{k,k+1} = s_{(k)} = \| \underline{v} \|_{2}$ & $7: \quad H_{k+1,k} = H_{k,k+1} = s_{(k)} = \| \underline{v} \|_{2}$\\
    $8: \quad \underline{q}_{(k+1)} = \underline{v}/s_{(k)}$ & $8: \quad \underline{q}_{(k+1)} = \underline{v}/s_{(k)}$ \\
  \otoprule
 \end{tabular}
 \caption{The Lanzcos algorithm and its physical counterpart which computes a tridiagonal matrix $H$ whose eigenvalues and eigenvectors are closely related to the eigenvalues and eigenvectors of $S_{11}^{H}\cdot S_{11}$. Note that we initialize the algorithm by setting $k=1$, $\underline{q}_{(1)}$ to a random unit norm vector, and $s_{(0)}=0$.}
\vspace{-0.45cm}
 \label{tab:ChanIdent}
\end{table}

\section{Numerical simulations and validation of the existence of highly transmitting eigen-wavefronts }
\label{sec:simulations}

\begin{figure}[B]
  \centering
  \includegraphics[width = 3.5in]{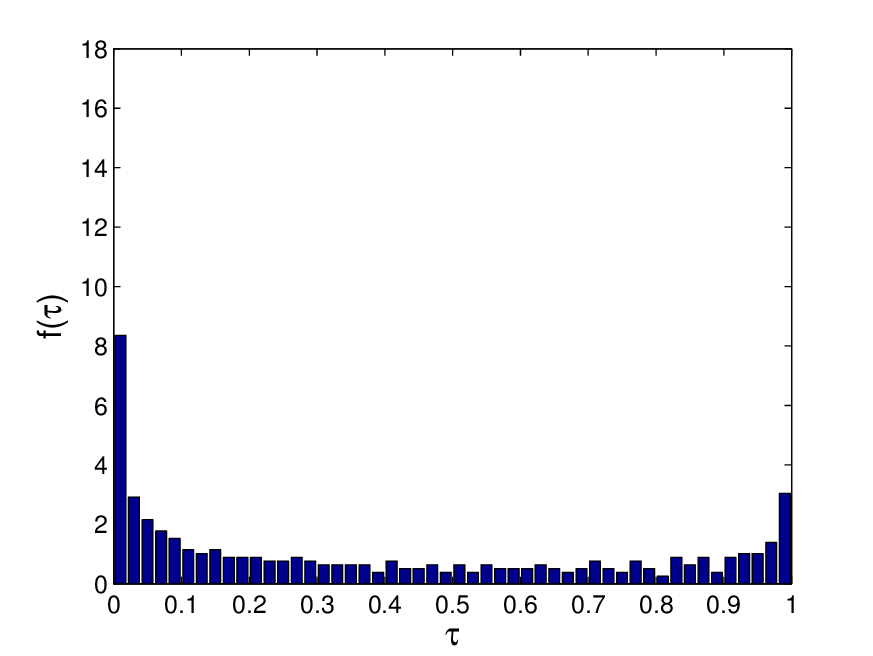}
  \caption{Empirical transmission coefficients distribution from a scattering system with $D=197\lambda, L=1.2\times10^{4}\lambda , r=0.11\lambda, N_{c} = 14,000$  (Dielectric) ,$ n_{d} = 1.3, M=395, \overline{l}=6.7\lambda$, where $\overline{l}$ is the mean of the minimum-inter-scatterer-distances.}
   \label{fig:dpmk numerical}
\end{figure}

\begin{figure}[!t]
\centering
\subfloat[Wavefield produced by a normally incident wavefront.]{\label{fig:open channel field norm}

\includegraphics[trim = 20 0 42 50, clip = true, height=3.35in]{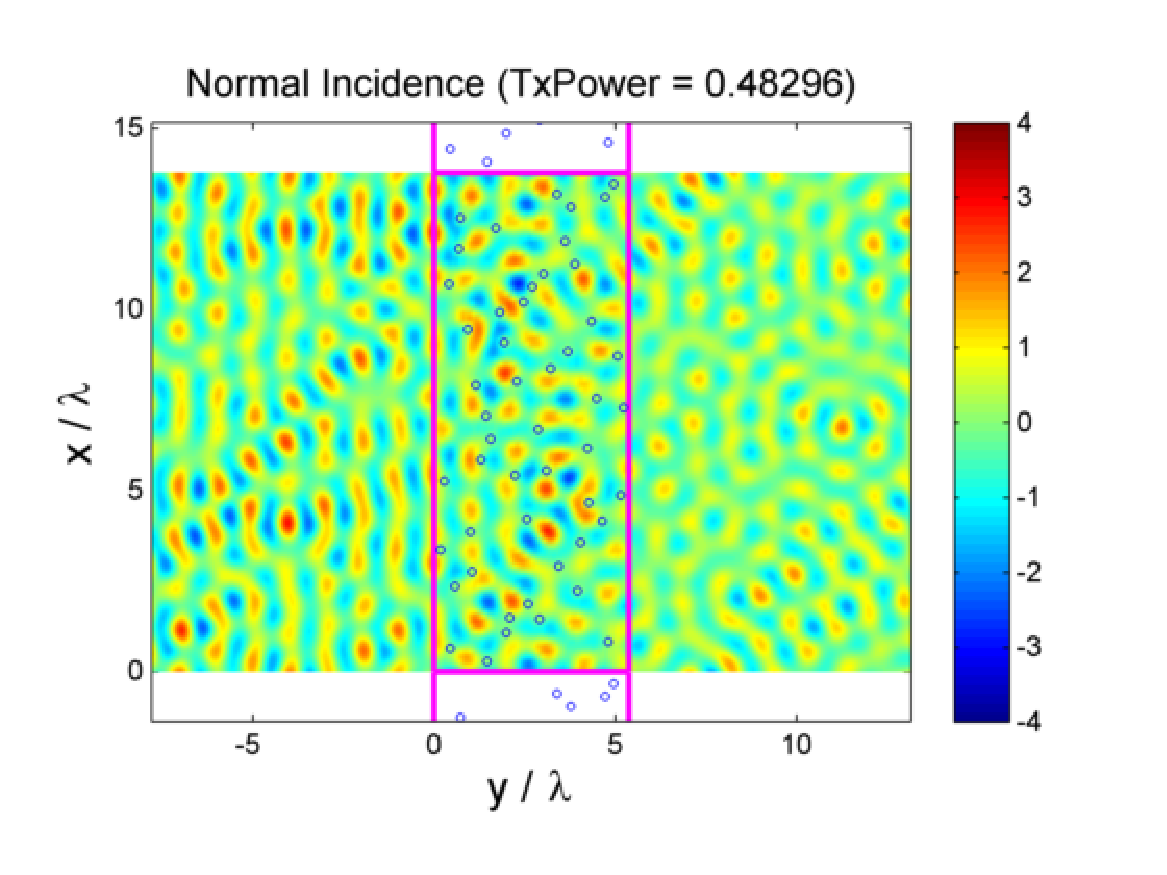}
}\\
\subfloat[Wavefield produced by the optimal wavefront.]{\label{fig:open channel field opt}

\includegraphics[trim = 20 0 42 50, clip = true,height=3.65in]{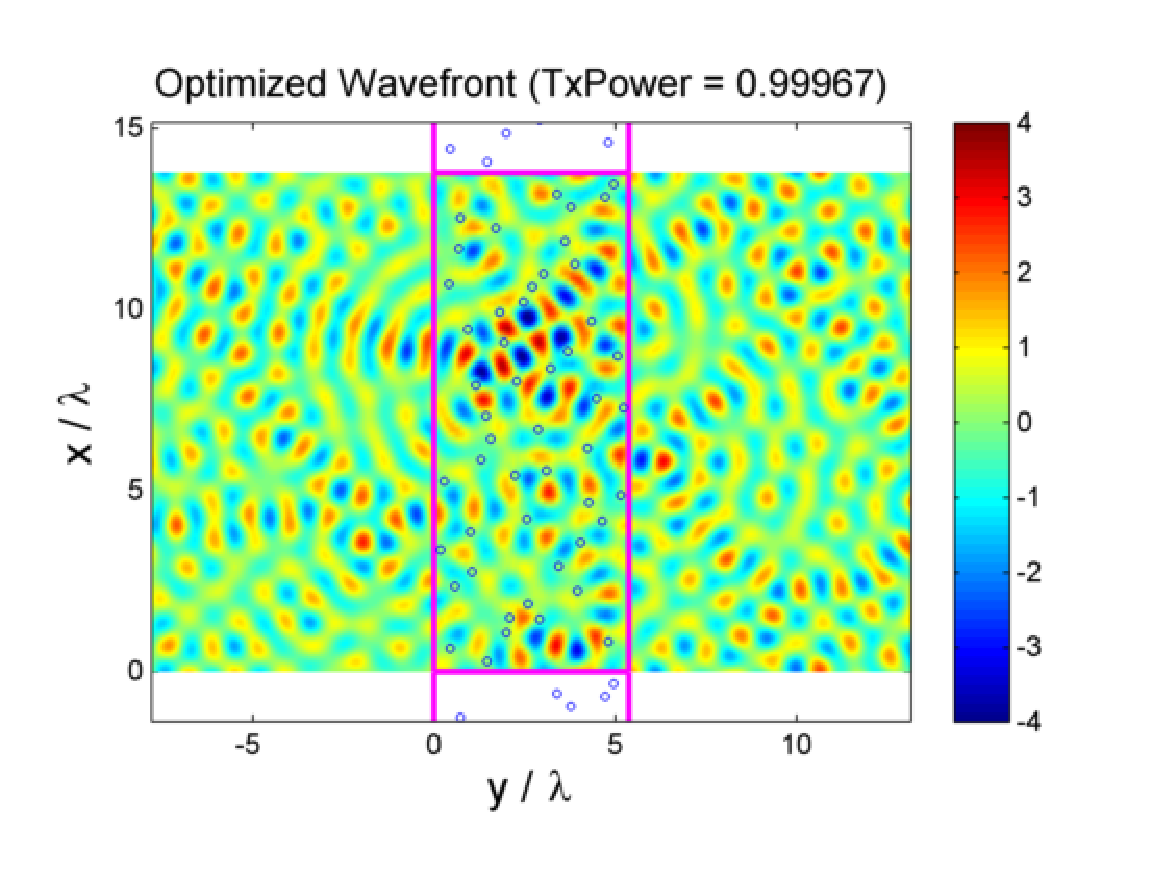}
}
\caption{Wavefield plot of the incident-plus-backscatter wave corresponding to (a) normally incident and the (b) optimal wavefront, which were sent to a scattering system with $D=14\lambda, L=5.4\lambda, r=0.11\lambda, N_{c} =50 \mbox{ PEC}, M=27, \overline{l}=0.8\lambda$. The normally incident wavefront has $\tnorm = 0.49$ while the optimal wavefront yields $\topt = 0.9995$.}
\label{fig:simulations}
\end{figure}

\begin{figure}
\centering
 \includegraphics[trim = 0 0 0 20, clip = true,height= 3.65in]{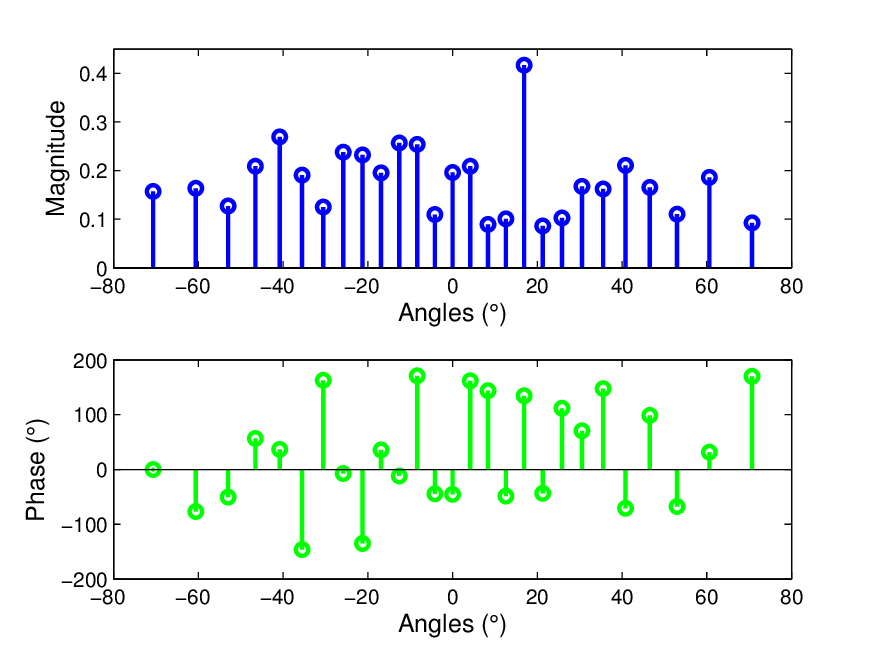}
\caption{The modal coefficients of the optimal wavefront corresponding to Fig \ref{fig:open channel field opt} are shown.}
\label{fig:opt coeff}
\end{figure}

\begin{figure}
\centering
\includegraphics[width=6.55in]{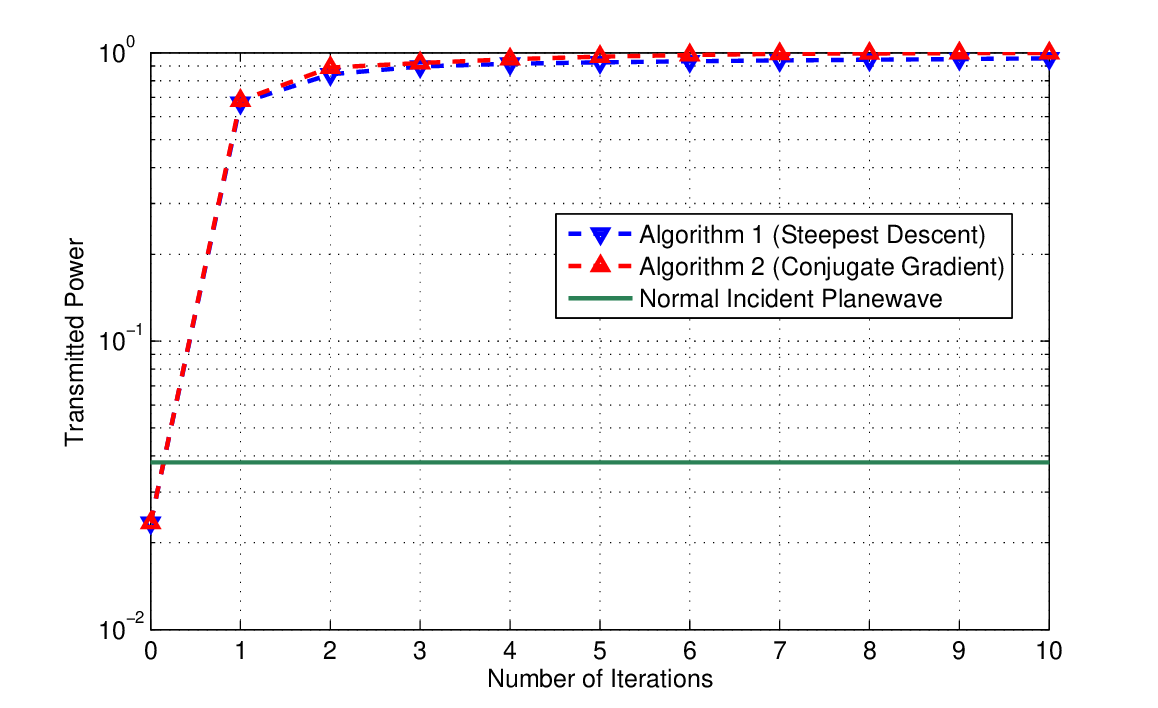}
\caption{The transmitted power versus the number of iterations is shown for steepest descent algorithm with $\mu =0.5037$ and for conjugate gradient in the setting with $D=197\lambda, L=3.4\times10^{5}\lambda, r=0.11\lambda, N_{c} = 430,000 \mbox{ dielectric cylinders with } n_{d} = 1.3, M = 395, \overline{l}=6.69\lambda$. The conjugate gradient algorithm converged to the optimal transmitted power slightly faster than the steepest descent algorithm. However, since the steepest descent algorithm requires a line search for setting the optimal step size $\mu$, it requires more measurements than the conjugate gradient method which does not require any parameters to be set.}\vspace{-0.05cm}
\label{fig:opt convergence}
\end{figure}

To validate the proposed algorithms, we compute the scattering matrices in Eq. (\ref{eq:scat matrix})  via a spectrally accurate, T-matrix inspired integral equation solver that characterizes fields scattered from each cylinder in terms of their traces expanded in series of azimuthal harmonics. Interactions between cylinders are modeled using 2D periodic Green’s functions. The method constitutes a generalization of that in \cite{mcphedran1999calculation}, in that it does not force cylinders in a unit cell to reside on a line but allows them to be freely distributed throughout the cell.  All periodic Green’s functions/lattice sums are rapidly evaluated using a recursive Shank's transform as in \cite{singh1991use,sidi2003practical}.  Our method exhibits exponential convergence in the number of azimuthal harmonics used in the description of the field scattered by each cylinder.  In the numerical experiments below, care was taken to ensure 11-th digit accuracy in the entries of the computed scattering matrices.

Fig.  \ref{fig:dpmk numerical} shows the empirical transmission coefficient distribution, \ie the singular value squared of the $S_{21}$ matrix of a slab with $D=197\lambda, L=1.2\times10^{4}\lambda , r=0.11\lambda, N_{c} = 14,000$  (Dielectric), $n_{d} = 1.3, M=395$ and , $\overline{l}=6.7\lambda$, where $\overline{l}$ is the mean of the minimum-inter-scatterer-distances. The computation validates the bimodal shape of the theoretical distribution in Fig. \ref{fig:dpmk}.

Next, we consider scattering system with $D=14\lambda, L=5.4\lambda, r=0.11\lambda, N_{c} =50$ (PEC), $M=27$, and  $\overline{l}=0.8\lambda$. Here $\tnorm = 0.49$ while $\topt = 0.9995$ so that wavefront optimization produces a two-fold increase in transmissited power.  Fig. \ref{fig:open channel field norm} and Fig. \ref{fig:open channel field opt} show the wavefield produced by the optimal wavefront and a normally incident wavefront, respectively.  Fig.  \ref{fig:opt coeff} shows the modal coefficients of the optimal wavefront corresponding to Fig. \ref{fig:open channel field opt}.

Fig.  \ref{fig:opt convergence} displays the rate of convergence of the algorithm's developed for a setting with $D=197\lambda, L=3.4\times10^{5}\lambda , r=0.11\lambda, N_{c} = 430,000$ (Dielectric), $n_{d} = 1.3, M = 395$ and, $\overline{l}=6.69\lambda$; this slab has a comparable (slightly lower) packing density than that in Fig.  \ref{fig:open channel field norm}.

A normally incident wavefront results in a transmission of $\tnorm = 0.038$. The optimal wavefront yields $\topt = 0.9973$  corresponding to a $26$-fold increase in transmission. Algorithms 1 and 2 produce wavefronts that converge to the near optimum in about $5-10$ iterations, as shown in Fig.  \ref{fig:opt convergence}.

\begin{figure}
\centering
\includegraphics[width=5.55in]{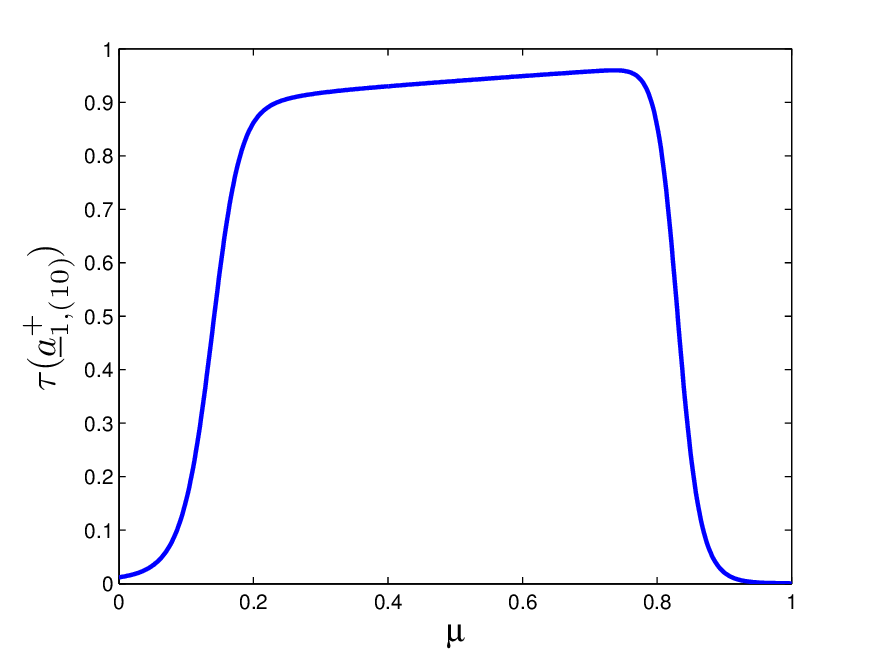}
\caption{The transmitted power at the $10$-th iteration as a function of the stepsize $\mu$ used in Algorithm 1 for the same setting as in Fig. \ref{fig:opt convergence}. }\vspace{-0.05cm}
\label{fig:MuSensitivity}
\end{figure}

Fig. \ref{fig:MuSensitivity} plots the transmitted power after the $10$-th iteration of Algorithm 1 for different choices of $\mu$.  Fig. \ref{fig:MuSensitivity} reveals that there is broad range of $\mu$ for which the algorithm converges in a handful of iterations. We have found that setting $\mu \approx 0.5$ yields fast convergence.

The conjugate gradient method (Algorithm 2) converges slightly faster than the steepest descent method (Algorithm 1) in the setting where we chose the optimal $\mu = 0.5037$ for Algorithm 1 by a line search; \ie we ran Algorithm 1 over a fixed set of discretized values of $\mu$ between 0 and 1, and chose the optimal $\mu$ that gives the fastest convergence result.  In an experimental setting, the line search for finding the optimal $\mu$ for the steepest descent algorithm will require additional measurements. Thus, Algorithm 2 will require fewer measurements than Algorithm 1 with the additional advantage of not requiring any auxiliary parameters to be set.

Next, we consider the setting where a subset of the propagation modes are controlled so that the summation in (\ref{eq:IncidentWave}) is from $-N_{\rm ctrl}$ to $N_{\rm ctrl}$. Thus the number of controlled modes is given by $M_{\rm ctrl} = 2 N_{\rm ctrl} + 1$.

\begin{figure}
\centering
\subfloat{
\includegraphics[width=6.05in]{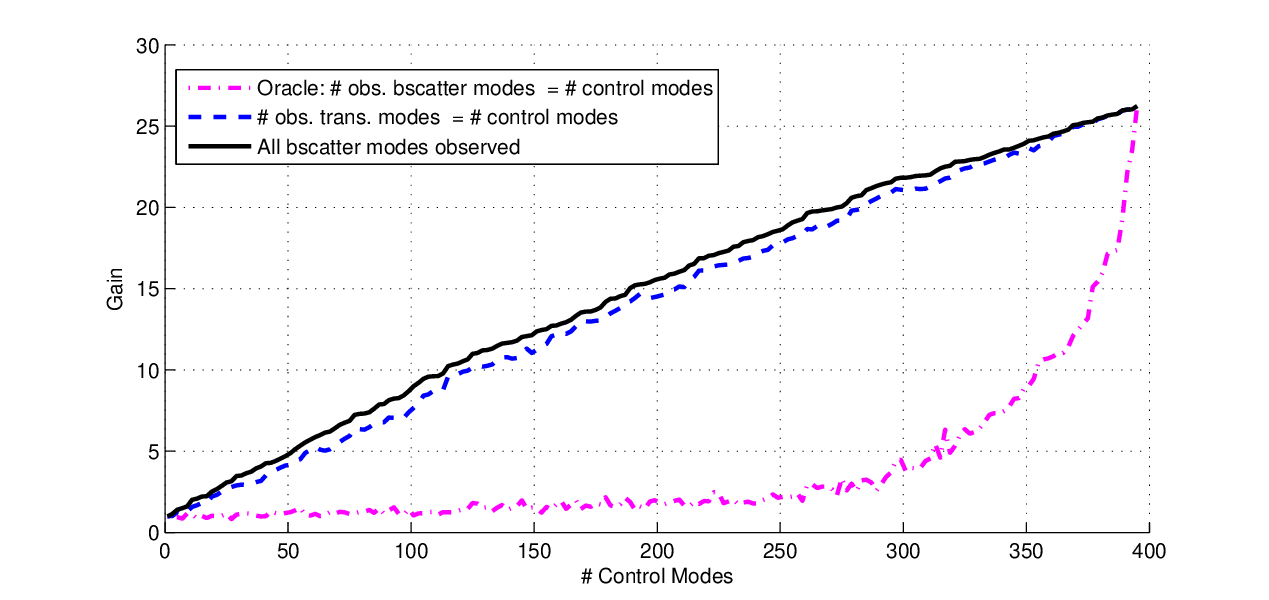}
}
\caption{Gain (=:$\topt / \tnorm$) versus the number of control modes for the same setting as in Fig. \ref{fig:opt convergence}. Here we compute the realized gain for algorithms that control only part of the total number of modes but capture, 1) all modes in the backscatter field, 2) only as many modes in the transmitted field as the number of control modes, and 3) only as many modes in the backscatter field as the number of control modes. For the last algorithm, we transmit the eigen-wavefront of the (portion of the) $S_{11}$ matrix that yields the highest transmission.}
\label{fig:GainDepAP}
\end{figure}

Fig. \ref{fig:GainDepAP} shows the realized gain (relative to a normally incident wavefront) for three different approaches versus the number of control modes in the same setting as in Fig.  \ref{fig:opt convergence}.   Here we compute the realized gain for algorithms that control only part of the total number of modes but capture, 1) all modes in the backscatter field, 2) only as many modes in the transmitted field as the number of control modes, and 3) only as many modes in the backscatter field as the number of control modes. For the last algorithm, we transmit the eigen-wavefront of the (portion of the) $S_{11}$ matrix that yields the highest transmission. Fig. \ref{fig:GainDepAP} shows that if the backscatter field is fully sampled, then it is possible to realize increased transmission with a limited number of control modes. It also emphasizes the important point that when the backscatter field is not fully sampled then the principle of  minimizing backscatter might produce `transmission' into the unsampled portion of the backscatter field instead of producing forward transmission.

\begin{figure}
\centering
\subfloat{
\includegraphics[width=5.8in]{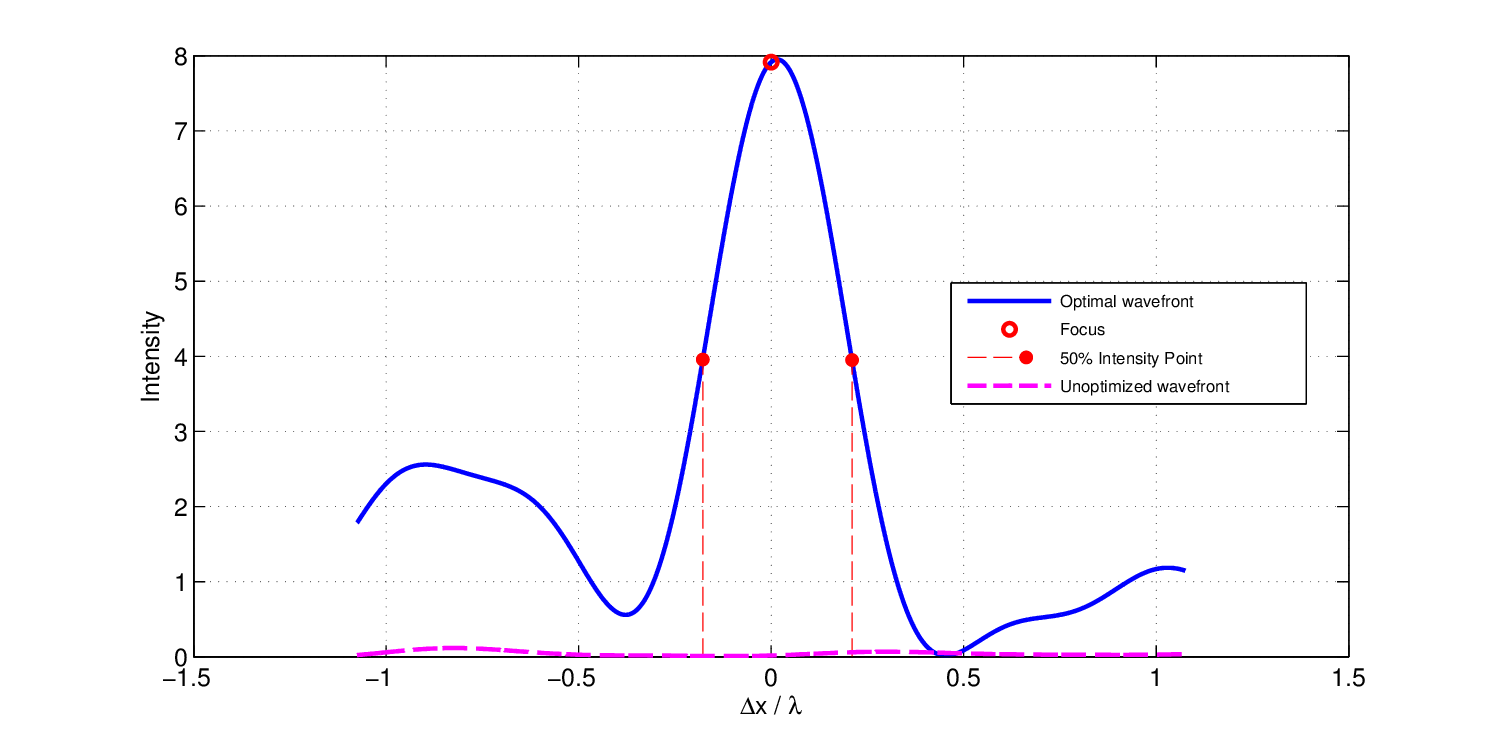}
}
\caption{Intensity plot around the target at $(D / 2,5.4\lambda)$ outside the scattering system defined in Fig.  \ref{fig:opt convergence}. The optimal focusing wavefront forms a sharp focus of  $1\lambda$ around the target. The unoptimized wavefront solution corresponds to an incident wavefront that would have produced a focus at the target if there were no intervening scattering medium.}
\label{fig:focus}
\end{figure}

\begin{figure}
\centering
\subfloat[]{\label{fig:IModeDecomp}

\includegraphics[width=4.5in]{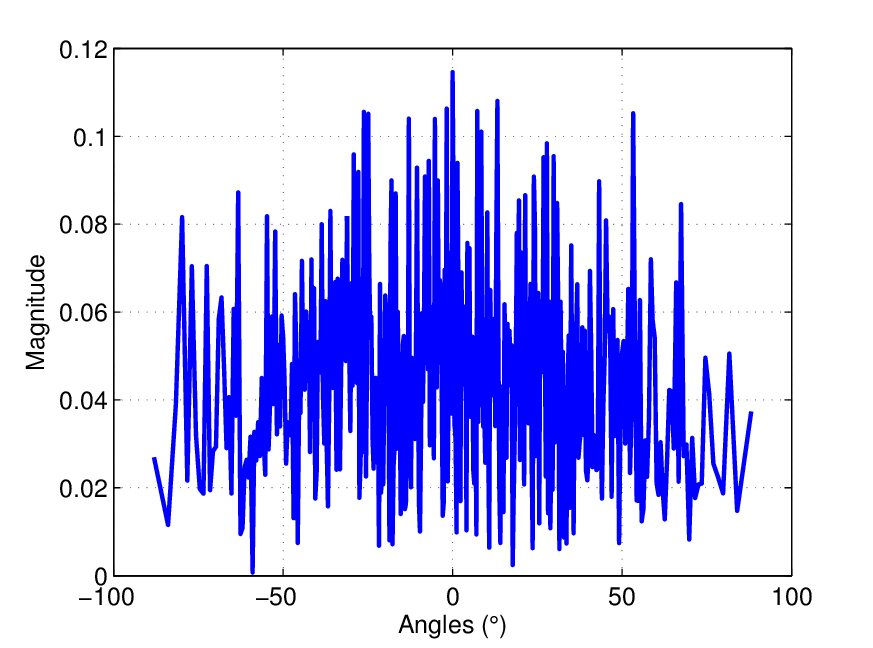}
}\\
\subfloat[]{
\includegraphics[width=4.5in]{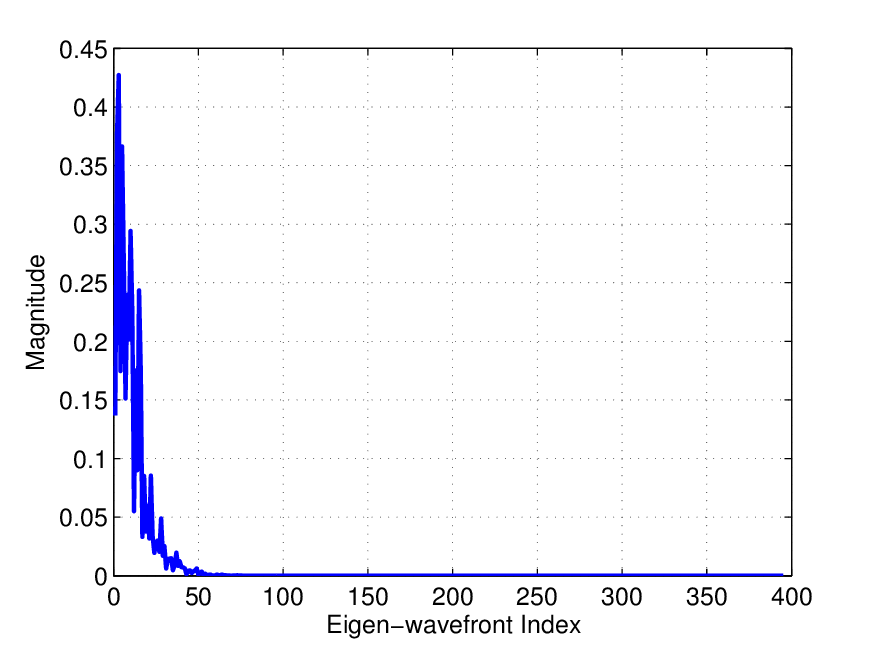}
\label{fig:EigModeDecomp}
}
\vspace{-0.65cm}
\caption{Here, we depict the magnitude of the coefficients of the optimal focusing wavefront, corresponding to the situation in Fig.  \ref{fig:focus}, in terms of two choices of bases vectors. In (a) we decompose the optimal focusing wavefront with respect to the bases vectors corresponding to plane waves; in (b) decompose the optimal focusing wavefront with respect to the bases vectors associated with the eigen-wavefronts of the $S_{11}$ matrix. A particular important observation is that the eigen-wavefront decomposition yields a sparse representation of the optimal focusing wavefront.}
\label{fig:modal coeff sparse}
\end{figure}

Fig.  \ref{fig:focus} considers the same setup as in Fig.  \ref{fig:opt convergence} with a target at $(D / 2,5.4\lambda)$ and plots the focus achieved at the target by exciting a focusing wavefront  as in (\ref{eq:afoc}).  The modal coefficients are plotted in Fig.  \ref{fig:IModeDecomp}. Fig.  \ref{fig:EigModeDecomp} shows the sparsity of the modal coefficients of the optimal focusing wavefront when expressed in terms of the basis given by the right singular vectors of the $S_{11}$ matrix or equivalently, the eigenvectors of $S_{11}^{H}\cdot S_{11}$.

\begin{figure}[!t]
\centering
\subfloat{
\includegraphics[width=6.35in,]{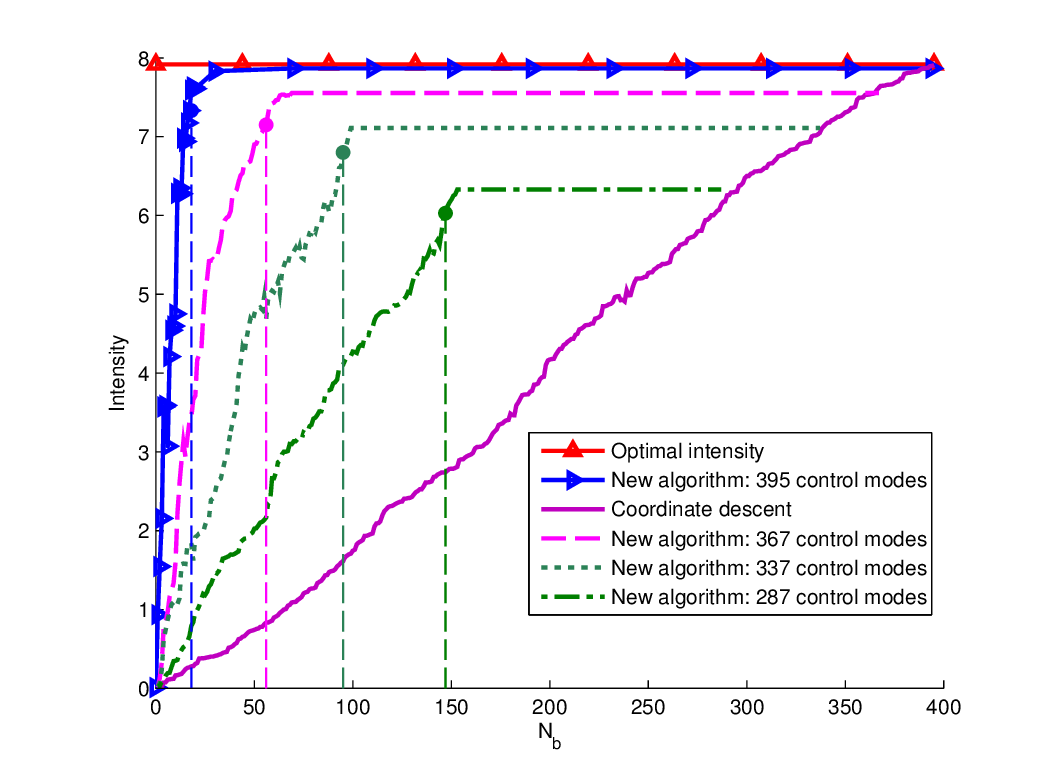}
}
\caption{Intensity at target as a function of the number of bases vectors for the new algorithm (which uses the bases vectors estimated using (\ref{eq:Bfoc}) and the  algorithm described in Table \ref{tab:ChanIdent}) for different number of control modes versus the standard coordinate descent method which uses the plane wave associated bases vectors (see Section \ref{sec:FocusingAlgorithms}) for the same setting as in Fig.  \ref{fig:focus}. The sparsity of the optimal wavefront's modal coefficient vector when expressed using the bases of the eigen-wavefronts (shown in Fig.  \ref{fig:EigModeDecomp})  leads to the rapid convergence observed. The optimal wavefront was constructed as described in Section \ref{subsec:Focusing} using time-reversal. The number of bases vectors needed to attain  $95$\% of the optimal focus intensity for a given number of control modes is indicated with a vertical line  highlighting the fast convergence of the algorithm and the ability to get a near-optimal focus using significantly fewer measurements than the coordinate descent approach.}
\label{fig:focus convergence}
\end{figure}

Fig.  \ref{fig:focus convergence} plots the intensity achieved when using $N_{B}$ bases vectors for the algorithms described in Section \ref{sec:FocusingAlgorithms} in the same setup as in Fig.  \ref{fig:focus}. The new algorithm which computes the bases $B$ from the eigenvectors of $S_{11}^{H}\cdot S_{11}$ associated with its smallest eigenvalues reaches $95$\% of the optimal intensity with significantly fewer iterations than the coordinate descent algorithm. This fast convergence to the near-optimum is the principal advantage of the proposed method. Figure \ref{fig:focus convergence} shows that this convergence behavior is retained even when the number of control modes is reduced. We obtain similar gains for the setting where there are multiple focusing points.

\begin{figure}
\centering
\subfloat{
\includegraphics[width=6.05in]{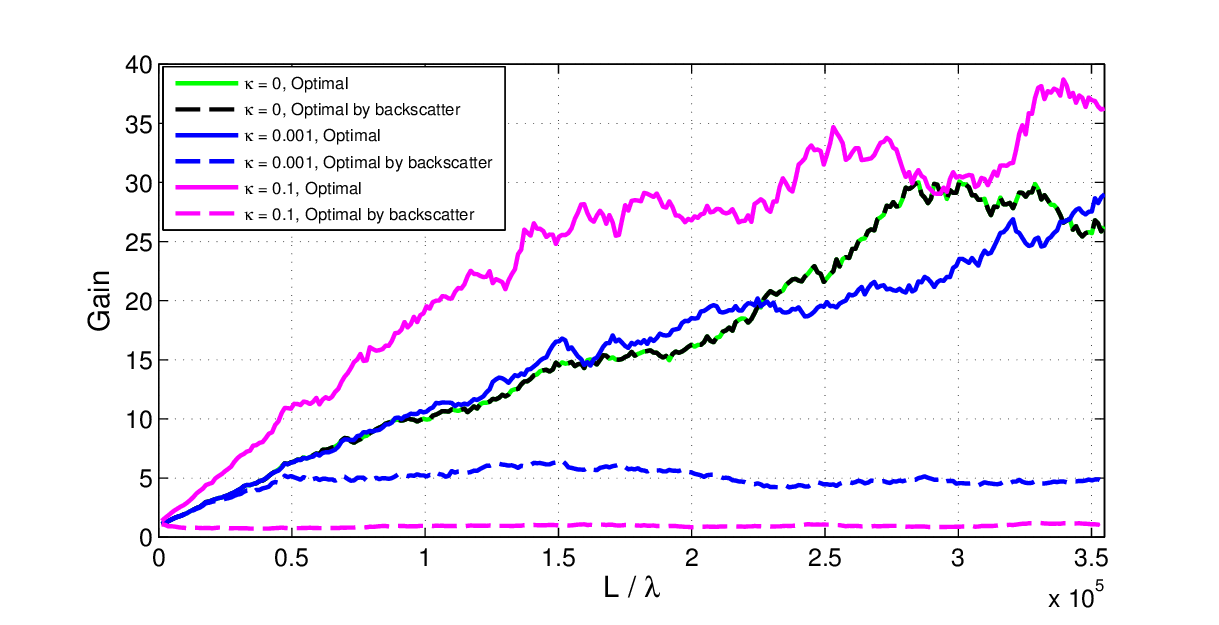}
}
\caption{Gain (=:$\topt / \tnorm$) versus the thickness $L/\lambda$ in a setting with $D=197\lambda, r=0.11\lambda, N_{c} = 430,000 \mbox{ Absorbing Dielectric}, n_{d} = 1.3-j\kappa, M = 395, \overline{l}=6.69\lambda$, for different values of  $\kappa$. The solid line represents the maximum possible gain and the dashed line represents the gain obtained by using backscatter minimizing algorithm discussed in Section \ref{sec:AlgoBackMin}.}
\label{fig:absorb}
\end{figure}

Finally, we consider the setting where the scatterers are absorptive. Here, backscatter minimization as a general principle for increasing transmission is clearly sub-optimal since an input with significant absorption can also minimize backscatter. We defined gain as $\topt /  \tnorm$. Here we have $D=197\lambda, L=3.4\times10^{5}\lambda, r=0.11\lambda, N_{c} = 4.3\times10^{5}$ (Absorbing Dielectric), $ n_{d} = 1.3-j\kappa, M = 395$, and  $\overline{l}=6.69\lambda$. In Fig.  \ref{fig:absorb}, we compare the gain obtained by using the backscatter minimizing wavefront to the gain obtained by the optimal wavefront (that utilizes information from the $S_{21}$ matrix) for various $\kappa$, as the thickness of the scattering system increases.  We obtain an increase in transmission and the methods described again produce dramatic gains whenever the scatterers are weakly absorptive.

\section{Conclusions}
\label{sec:conclusions}

We have numerically verified the existence of eigen-wavefronts with transmission coefficients approaching one in highly scattering systems and developed physically realizable algorithms for finding these highly transmitting eigen-wavefronts using backscatter analysis. We also developed a physically realizable algorithm for forming a focused input using the highly transmitting eigen-wavefronts identified by the previous algorithm. Via numerical simulations it was shown that the algorithms converged to a near-optimal wavefront in just a few iterations. The proposed algorithms are quite general and may be applied to scattering problems beyond the 2-D setup described in the simulations. We are currently investigating extensions to imaging and sensing applications. A detailed study of the impact of periodic boundary conditions on the results obtained is also underway. 

\section*{Acknowledgements}

This work was partially supported by NSF grant CCF-1116115, an AFOSR Young Investigator Award FA9550-12-1-0266 and an AFOSR DURIP Award FA9550-12-1-0016.

\vspace{-0.25cm}

\appendix




\end{document}